\documentclass[a4paper,11pt]{article}
\usepackage[pdftex]{graphicx}
\usepackage[T1]{fontenc}
\usepackage{lmodern}
\usepackage{slashed}
\usepackage[utf8]{inputenc}
\usepackage[english]{babel}
\usepackage{microtype}
\usepackage{cite}
\usepackage{amsmath}
\usepackage{amssymb}
\usepackage{amsfonts}
\usepackage[nottoc,notlot,notlof]{tocbibind}
\usepackage{upgreek}
\usepackage{mathtools}
\numberwithin{equation}{section}
\allowdisplaybreaks
\usepackage[all]{xy}
\usepackage{color}
\usepackage{graphicx}
\graphicspath{{images/}}
\usepackage{geometry}
\usepackage[toc,page]{appendix}
\usepackage{hyperref}
\usepackage[normalem]{ulem}

\newcommand{\diff}{\mathrm{d}}

\newcommand{\ii}{\mathrm{i}}

\usepackage{bm}
\usepackage{ragged2e}
\usepackage{appendix}
\usepackage{slashed}
\usepackage{bbold}
\usepackage{cancel}

\definecolor{blue-violet}{rgb}{0.54, 0.17, 0.89}
\definecolor{PineGreen}{cmyk}{0.92, 0, 0.59, 0.25}
\definecolor{YellowOrange}{cmyk}{0, 0.42, 1, 0}
\definecolor{orange}{rgb}{0.95, 0.5, 0.1}

\DeclareMathAlphabet{\mathpzc}{OT1}{pzc}{m}{it}
\newcommand{\virgolette}{``}

\usepackage[dvipsnames]{xcolor}

\begin{document}

\begin{titlepage}
\begin{flushright}
\par\end{flushright}
\vskip 0.5cm
\begin{center}
\textbf{\LARGE \bf M5-brane in the superspace approach}\\
\vskip 5mm

\vskip 1cm

\large {\bf L.~Andrianopoli}$^{~a, b}$\footnote{laura.andrianopoli@polito.it},
\large {\bf C.~A.~Cremonini}$^{~c}$\footnote{carlo.alberto.cremonini@gmail.com}, 
\large {\bf R.~D'Auria}$^{~a, b}$\footnote{riccardo.dauria@polito.it},
\large {\bf P.~A.~Grassi}$^{~b, e}$\footnote{pietro.grassi@uniupo.it},
\large {\bf R.~Matrecano}$^{~a, b}$\footnote{riccardo.matrecano@polito.it},
\large {\bf R.~Noris}$^{~d}$\footnote{noris@fzu.cz},
\large {\bf L.~Ravera}$^{~a, b}$\footnote{lucrezia.ravera@polito.it},
\large {\bf M.~Trigiante}$^{~a,b}$\footnote{mario.trigiante@polito.it}

\vskip .5cm {
\small
{
$^{(a)}$
\it Politecnico di Torino, Corso Duca degli Abruzzi, 24, 10129 Torino, Italy}\\
{$^{(b)}$ \it INFN, Sezione di Torino}
{\it via P.~Giuria 1, 10125 Torino, Italy} \\
{$^{(c)}$ \it Faculty of Mathematics and Physics, Mathematical Institute, Charles University Prague, Sokolovska 49/83, 186 75 Prague}\\
{$^{(d)}$ \it CEICO, Institute of Physics of the Czech Academy of Sciences, \\
Na Slovance 2, 182 21 Prague 8, Czech Republic}\\
{$^{(e)}$
\it DiSIT,} 
{\it Universit\`a del Piemonte Orientale,} 
{\it viale T.~Michel, 11, 15121 Alessandria, Italy}
}
\end{center}

\begin{abstract} 
{Motivated by Sen's spacetime prescription for the construction of theories with self-dual field strengths,} we {present} a rigid superspace Lagrangian describing noninteracting tensor multiplets living on a stack of M5-branes and containing all the physical constraints on the fields, yielding the on-shell matching of {the} degrees of freedom. {The geometric superspace approach adopted here offers a natural realization of superdiffeomorphisms and is particularly well suited for the coupling to supergravity. However, within this formulation the} {(anti-)}self-duality property of the $3$-form field strengths {is lost} when the {superspace} Lagrangian is trivially restricted {to} spacetime. We {propose} two main paths to {address} this issue: {a first-order superspace extension of Sen's spacetime results, which, once trivially restricted to spacetime, yields all the dynamical equations including the (anti-)self-duality constraint on the $3$-form field strengths, and a possible way to obtain a full superspace description of the theory, based on integral forms.}

\end{abstract}

\vfill{}
\vspace{1.5cm}
\end{titlepage}

\setcounter{footnote}{0}
\tableofcontents

\section{Introduction}

A long-standing problem in QFT and supergravity is the construction of theories with self-dual field strengths. Those theories are ubiquitous and, although several studies have been carried out, a completely satisfactory formulation is still missing. \\
The main problem can be summarized by the following question: How does one define a consistent variational principle such that the corresponding equations reproduce the Euler-Lagrange equations of motion comprehensive of the self-duality constraints? \\ 
This problem already appears in theories involving only bosonic degrees of freedom (d.o.f.): in any ${(4n+2)}$-dimensional model, one can consider ${(2n)}$-form potentials ${A^{(2n)}}$ whose associated field strengths ${F^{(2n+1)}}$ are self-dual or anti-self-dual ${(2n+1)}$-forms $F^{(2n+1)}=\pm \star F^{(2n+1)}$ with respect to a given Hodge dual operator $\star$ defined on the $(4n+2)$-dimensional (pseudo-)Riemannian manifold. 
{This issue becomes particularly relevant for} chiral supersymmetric theories in $(4n+2)$ spacetime dimensions, {where the (anti-)self-dual field strengths are real, and} the self-duality constraint is required for the matching of the on-shell degrees of freedom {implied by} supersymmetry (susy). \\ 
{In all of these theories, however, a Lagrangian formulation is problematic, since the kinetic term of self-dual field strengths in $D=(4n+2)$ dimensions vanishes.
An example of such an issue is given by} the tensor multiplet {in the chiral $N=(4,0)$ (16 supercharges) theory in} six dimensions, which describes the world volume theory of a single M5-brane. {The multiplet contains} a spinor {$\lambda_A$} which -- on-shell -- propagates only 8 real degrees of freedom (the Dirac equation halving the spinorial degrees of freedom){. The odd degrees of freedom are} paired with {a bosonic field content given by} five {real} scalars $\phi^{[AB]_0}$ and a $2$-form $B^{(2)}$, whose field strength $H^{(3)}$ is (anti-)self-dual, {thus} carrying 3 real d.o.f. {and allowing the} matching of the fermionic ones \cite{Bergshoeff:1987cm,Polchinski:1995mt,Witten:1995em,Townsend:1995af,Duff:1990wv}.\\
{A Lagrangian description of these theories using unconstrained off-shell fields, and implementing the self-duality constraint on-shell among the Euler-Lagrange constraints would be desired.}

However, the off-shell matching of degrees of freedom in supersymmetric theories is in general problematic for {theories with eight supercharges or more, that is for extended supergravities in four dimensions and} for  higher-dimensional models (as in the case M5-brane in $D = 6$ and higher).\footnote{Notice that sometimes the two problems, absence of auxiliary fields and self-duality constraints, are the two faces of the same medal. For example in the case of $D=4$ $N=4$ super-Yang-Mills theory, the equations of motion are implemented by requiring a self-dual condition in the $R$-symmetry indices of the scalar superfields. The difficulties to implement this  constraint as a variational principle are equivalent to the self-dual field strengths.}
This makes the construction of an action principle for path integral computations (for example in localization methods) {a very} difficult task.

In the past, there have been several attempts to circumvent these problems (see for example the pioneering works 
\cite{AlvarezGaume:1983ig,Henneaux:1988gg,Bastianelli:1989hi}), using  different techniques such as non-Lorentz-covariant formulation, infinite number of auxiliary fields, and nonpolynomial actions (see for example \cite{Floreanini:1987as,Bastianelli:1989cu,McClain:1990sx,Pasti:1996vs,Berkovits:1996nq,Berkovits:1996em,Howe:2000nq, Devecchi:1996cp,Pasti:1997gx,Schwarz:1997mc,DallAgata:1997gnw,DallAgata:1998ahf,Castellani:2019qmb,Mkrtchyan:2019opf}), each of which has its own advantages and drawbacks.
{Among them, it is worth mentioning the geometric superspace approach developed in \cite{cube}, where the self-duality constraint can be obtained on-shell from a superspace Lagrangian. This was applied in particular in   \cite{DAuria:1983jkr, Castellani:1993ye}. However, in this approach the self-duality constraint emerges when analyzing the Euler-Lagrange equations in the whole superspace, while the restriction to spacetime of the superspace Lagrangian fails to be invariant and to yield, among the field equations on spacetime, the self-duality constraint.}
Nonetheless, a completely satisfactory formulation was not available, until the recent works by Sen \cite{Sen:2015nph,Sen:2019qit}, based on string field theory, rejuvenating the field and prompting new developments \cite{Lambert:2019khh,Lambert:2019diy,Townsend:2019koy,Bandos:2020hgy,Buratti:2019guq,Andriolo:2020ykk,Bansal:2021bis,Avetisyan:2022zza,Vanichchapongjaroen:2020wza,Gustavsson:2020ugb,Andriolo:2021gen,Arvidsson:2004xa,1837177}. A preliminary remark is in order: although Sen's formulation avoids all problematic features of previous approaches, it has to deal with a nonconventional realization of superdiffeomorphisms. This is justified by the string field theory approach, but the analysis has been pursued only in the component formalism. 
In addition, we have to recall that the derivation discussed in \cite{Sen:2015nph,Sen:2019qit} has only been {carried out} in weak gravity approximation on a flat background and that a complete supergravity analysis is still missing. \\

In the present work, we provide a superspace Lagrangian whose Euler-Lagrange equations {in superspace} include the self-duality constraint on the $3$-form field strength and whose restriction to spacetime, setting $\theta=0=\diff \theta$, is globally invariant under supersymmetry, describing at lowest order the world volume theory of the M5-brane. On the other hand, the obtained theory can be considered as a testing ground, where to advance proposals for superspace prescriptions, implementing the self-duality constraint directly on spacetime, which will then be tested in future works in cases of local supersymmetry.\\

Historically, there have been two ways to describe supersymmetric theories or supergravities using a superspace approach: a first Lagrangian method based on superfields and superderivatives {(see \cite{gates})} and a second method based on the geometry of supermanifolds (see \cite{cube}). 
The latter is a powerful framework for the formulation of supergravity and rigid supersymmetric theories, often referred to as the \emph{geometric}, or \emph{rheonomic}, approach. It has proven to be a valuable asset in the construction of supersymmetric theories in various dimensions and degrees of supersymmetry, providing a consistent formulation also in certain cases where a spacetime action description was not available. In this formalism, the full local symmetry structure of the theory, including its supersymmetric properties, is encoded in the formal definition of the superfield strengths and their constrained parametrizations, which consist in their expansion on a basis of the cotangent bundle of superspace, generated by the vielbein $V^a$ and the gravitino $\psi^A$ 1-superforms. The consistency between these parametrizations and the Bianchi identities satisfied by the set of field strengths, yields a number of constraints on the superfields of the theory. These data encode, in an intrinsically geometric fashion, the supersymmetry transformation rules and their closure on the fields of the model, modulo local symmetry transformations. They also yield dynamical equations and all other constraints, including the (anti-)self-duality property of the chiral forms, allowing for the on-shell matching of degrees of freedom. \\
The geometric approach has a further outcome, which is the construction of a $D$-superform Lagrangian,\footnote{This Lagrangian is a bosonic $D$-superform, which can be integrated over a bosonic $D$-dimensional hypersurface in superspace, defining spacetime.} on the $\mathcal{M}^{(D|N)}$ superspace, whose Euler-Lagrange equations reproduce the aforementioned constraints on the fields, independently derived from the closure of the Bianchi identities. More precisely, the same equations, restricted to spacetime, yield the dynamical field equations, while their components along the other directions of superspace encode further information on the theory, related to the closure of supersymmetry transformations on the local symmetries of the model. As we shall see in the in the present work, and as shown in earlier analyses, in the chiral models under consideration in $D=4n+2$,  the (anti-)self-duality condition on the field strengths of the $2n$-forms potentials, is enforced by components of the Euler-Lagrange superspace constraints along odd directions. As a consequence of this, the same conditions, which represent the field equations for the $2n$-forms, do not follow from a spacetime action principle, provided the spacetime Lagrangian is defined through the trivial restriction of the superspace to spacetime, effected by setting $\theta=0$ and $\diff \theta=0$. One of the aims of the present analysis is to 
discuss this seeming drawback of the geometric formulation, in the specific rigid toy model under consideration, and to suggest possible equivalent definitions of the super-Lagrangian, which yield, once restricted to spacetime, Sen's construction. This would provide a simple superspace extension of the latter, paving the way for the interacting and supergravity cases.

The two superspace approaches can be successfully reformulated into a single framework of the {\it integral} forms approach \cite{Witten:2012bg,Castellani:2014goa,Castellani:2015paa,Grassi:2016apf,Castellani:2015ata,Castellani:2017fhi}. Given the 
rheonomic ${D}$-form Lagrangian ${\mathcal L}^{(D)}(\Phi, \diff\Phi, V, \psi)$ written in terms of the fields $\Phi$, of their differentials $\diff\Phi$ and of the supervielbein ($V^a, \psi^A$), one can build an action by integrating it on the entire supermanifold ${\mathcal M}^{(D|N)}$, to be indentified in this case with the worldvolume of the M5-brane. This requires the integrand to be an {\it integrable} form \cite{Witten:2012bg,Catenacci:2010cs,Catenacci:2018xsv}, which can be achieved by representing the embedding of a bosonic $D$-dimensional submanifold into the supermanifold ${\mathcal M}^{(D|N)}$ using the Poincar\'e dual form $\mathbb{Y}^{(0|N)}$ (where the second superscript denotes the {\it picture number} \cite{Catenacci:2010cs,Castellani:2014goa}, which must match the fermionic dimension of the 
supermanifold). The integrable form to be integrated is now ${\mathcal L}^{(D)}(\Phi, \diff \Phi, V, \psi)\wedge \mathbb{Y}^{(0|N)}$ and gives rise to a proper action, suitable for the variational derivation of the equations of motion. By changing the embedding, $\mathbb{Y}^{(0|N)}$ changes 
by exact terms $ \mathbb{Y}^{(0|N)} + \diff \Sigma$ which are harmless if the Lagrangian is closed $\diff  {\mathcal L}^{(D)}(\Phi, \diff \Phi, V, \psi)=0$. {In that case, the Euler-Lagrange equations derived without considering $\mathbb{Y}^{(0|N)}$ coincide with the equations arising from the variation of the action for \emph{any} choice of the embedding described. This means that any choice of $\mathbb{Y}^{(0|N)}$ gives rise to the same equations of motion, but with different manifest symmetries.} \\
This is, however, not possible in the presently considered case of six-dimensional tensor multiplets, and in theories without auxiliary fields for off-shell supersymmetry, where the rheonomic Lagrangian ${\mathcal L}^{(D)}(\Phi, \diff \Phi, V, \psi)$ fails to be closed. This means that the Poincar\'e dual form cannot be ignored and will project out some of the equations, as it happens in the $\theta=0=\diff \theta$ case. More general embeddings have been considered in \cite{Cremonini:2020skt}, where two of the authors of the present work proposed a method for writing an action, starting from the geometric Lagrangian for the supersymmetric chiral boson. We will discuss, inspired from that result, the possible generalization of such procedure to the case considered here, which will possibly make use of the superspace Hodge dual operator defined in \cite{CCGir,Castellani:2015ata,Cremonini:2020skt}.
This will be the object of a forthcoming publication.

{As a concluding remark, let us add that} the extension of Sen's approach in the presence of gravity, though valuable, requires a rather involved derivation that appears somewhat more contrived than in the rigid case. 
One of the motivations of the present analysis is a superspace generalization of Sen's mechanism in presence of gravity, which  will be left to future endeavors.

The paper is organized as follows: In Sec. \ref{Sec2} we review the fundamental concepts of the geometric approach, which will be used in Sec. \ref{Sec3}, where we will introduce the dynamical fields and perform the preliminary analysis of the Bianchi identities, identifying the constraints that the chosen fields have to satisfy on-shell {and their supersymmetry transformations}. In the same section, we will also present the $\mathcal L^{(6|0)}$ Lagrangian and discuss its features and its trivial projection on spacetime. In Sec. \ref{Sec4} we will {introduce a first prescription for modifying the geometric Lagrangian, which yields Sen's prescription, when trivially restricted to spacetime. In Sec. \ref{Towards full description with integral forms} we will instead focus on alternative ways of dealing with this problem, by considering nonfactorized integral form Lagrangians and nontrivial projections on spacetime.} 

\section{Lagrangian, Action and Supersymmetry}\label{Sec2}

In this short section, we review relevant aspects of the geometric approach to supegravity.

\subsection{Rheonomy in a nutshell}\label{rheonut}
Usually there is a twofold way to obtain a geometric formulation of the theory without using coordinates in superspace but using only $p$-forms:
\begin{itemize}
\item
an action principle formulated in a nonstandard way, since the Lagrangian is not integrated  on the full supermanifold ${\mathcal M}^{(D|N)}$, but only on a $D$-dimensional hypersurface embedded in superspace;

\item
a purely algebraic method based on the Bianchi identities of the super-field strength $2$-forms (to be referred also to as supercurvature $2$-forms)  as derived from the Maurer-Cartan equations  of a Lie superalgebra (or $p$-forms supercurvatures derived from a free differential algebra).
\end{itemize}
In the latter case, one writes down expressions of the curvature $p$-forms expanded along the $p$-dimensional basis of supercotangent bundle (given by exterior products of the bosonic and fermionic vielbein), which have to be compatible with all the symmetries of the theory (Lorentz invariance, scaling behavior, etc.). One then assumes the following  requirement: all  the components of the curvatures along a basis featuring at least one fermionic vielbein $\psi^\alpha$ should be expressed in terms of the supercurvature components along the bosonic vielbein $V^{a_1}\wedge V^{a_2} \wedge \dots \wedge V^{a_p}$. These latter components only have antisymmetric rigid bosonic indices and, once expressed in terms of the spacetime differentials (holonomic dual basis), are actually the so-called supercovariant field strengths in the Noether approach.

Such a requirement is called \emph{rhenomy principle} and allows one to not introduce extra degrees of freedom in the theory besides the physical ones. By requiring the closure of the Bianchi identities of the parametrized curvatures, one fixes the constant coefficients left undetermined. However, in all the theories where the number of bosonic and fermionic degrees of freedom only matches on-shell, the closure of the Bianchi identities also requires differential constraints on the supercovariant field strengths, which are nothing else than the equations of motion. Besides, one often finds further constraints which cannot be seen in a purely spacetime approach. Moreover, since the susy transformations are Lie derivatives in superspace, which, using the anholonomic parameter $\epsilon^\alpha$, can be written in terms of the gauge transformations plus contraction of the curvature terms, it is clear that the knowledge of the given parametrization of the curvatures also determines the susy transformations of the fields.\\

{In the former case, the Lagrangian depends generically on the superfields with the obvious constraint of respecting all the symmetries of the theories and, most importantly, it is a $D$-form, $D$ being the dimension of the bosonic hypersurface $\mathcal{M}^{(D)}$ of integration (representing spacetime), immersed in superspace. This implies that the Euler-Lagrange equations obtained by the variation of the $D$-form Lagrangian, which generically are $k$-form equations, with $k\leq D$, can be extended to the full superspace and can be analyzed along all the basis elements of the $D$-dimensional cotangent space spanned by different combinations of bosonic and fermionic vielbein $V^k,V^{k-1}\psi,\dots, \psi^k$. 
In order for the Lagrangian $D$-form to be independent of the embedding of $\mathcal{M}^{(D)}$ in superspace, it must be constructed only in terms of differential forms, exterior derivatives and wedge products, without using the spacetime Hodge operator. As a consequence, the bosonic kinetic terms should be written in a first-order formalism, by introducing suitable 0-forms auxiliary fields.}

It turns out that the analysis of the equations of motion along $V^k$ gives dynamical equations for the supercovariant field strengths, which must and do coincide with those obtained from the Bianchi identities. By projecting these equations along the $\diff x^{\mu_1}\wedge \diff x^{\mu_2}\wedge \dots \wedge \diff x^{\mu_k}$ $k$-forms, one recovers the spacetime equations.\\
The analysis of the equations of motion obtained along any basis featuring at least one $\psi^\alpha$ gives instead linear relations expressing the supercurvatures with one or more ``legs'' along $\psi$ in terms of the supercovariant field strengths along $V^k$. These are precisely the rheonomic conditions required in the Bianchi identities approach. Therefore they are not to be imposed, but come out {as a consequence of the Euler-Lagrange equations}. Moreover one can also often obtain further algebraic constraints on the supercovariant curvatures that are not visible in a purely spacetime approach.
Actually, the best way to construct this geometrical approach is to make use of the parametrization of the curvatures in order to simplify the analysis of the equations of motion of the Lagrangian.

Finally we observe that the invariance of the Lagrangian under supersymmetry is already built in using the geometrical approach: indeed if one performs the Lie derivative $\iota_\epsilon \diff +\diff \iota_\epsilon$ along a supersymmetry tangent vector $\overrightarrow{\nabla}=\epsilon^\alpha \nabla_{\alpha}$, discarding the total derivative $\diff \iota_\epsilon \mathcal{L}$, one obtains that the contraction on the $\psi$ fields gives $\epsilon$, while the contraction of the curvatures gives constraints on them which coincide exactly with the rheonomic constraints as obtained from the Lagrangian. This makes $\iota_\epsilon \diff \mathcal{L}=0$ identically, so that the Lagrangian is invariant in all superspace (that is even if evaluated on other hypersurface) and in particular on spacetime.

\subsection{Extension to the full superspace}
{To formulate a well defined action principle in superspace, it is desirable to extend the bosonic $D$-form Lagrangian discussed above to a $(D|N)$-form to be integrated over the full supermanifold  ${\mathcal M}^{(D|N)}$, where $N$ is the fermionic dimension. This requires using the integral-form formalism introduced by some of the authors in {\cite{Castellani:2014goa,Castellani:2015paa}} and whose main ingredients are summarized in Appendix \ref{integral}. }
To this end, we rename by ${\mathcal L}^{(D|0)}(\Phi, \diff \Phi, V, \psi)$ the {$D$-form} Lagrangian in superspace constructed along the lines discussed above, and previously referred to as $\mathcal{L}^{(D)}$. It is a $(D|0)$-form depending on the dynamical fields of the theory $\Phi$, their differentials $\diff\Phi$ and on the supervielbein $(V^a, \psi^A)$, whose dynamics will not be addressed in this paper. {To perform }the embedding of {the bosonic submanifold} ${\mathcal M}^{(D)}$ into $\mathcal{M}^{(D|N)}$, we first introduce the 
{super-}Poincar\'e dual $\mathbb{Y}^{(0|N)}$: it is a nontrivial cocycle in $\mathcal{M}^{(D|N)}$, and any variation of the embedding corresponds to a trivial deformation, belonging to the same cohomology class: 
\begin{eqnarray}
\label{PCOA}
\diff \mathbb{Y}^{(0|N)} =0\,, ~~~~~~~
\mathbb{Y}^{(0|N)} \neq \diff \Sigma^{(-1|N)}\,,~~~~~~~
\delta \mathbb{Y}^{(0|N)} = \diff \Gamma^{(-1|N)}\,. 
\end{eqnarray}
Notice that sometimes one can choose $\mathbb{Y}^{(0|N)}$ to respect some symmetries manifestly: $\delta \mathbb{Y}^{(0|N)} =0$. The details of the structure of  $\mathbb{Y}^{(0|N)}$ are discussed in Sec. \ref{Towards full description with integral forms} and in Appendix \ref{integral}. Further details can be found in the literature \cite{manin}.\\  
The forms $\Sigma^{(-1|N)}$ and $\Gamma^{(-1|N)}$ are $(-1|N)$ forms which can be written in terms of derivatives of Dirac deltas $\delta(\diff \theta)$. 
{Requiring t}he vanishing {of} a generic variation of the Lagrangian, $\delta {\mathcal L}^{(D|0)}(\Phi, \diff \Phi, V, \psi) =0$, implies the Euler-Lagrange equations of motion. In addition, we note that, since 
${\mathcal L}^{(D|0)}(\Phi, \diff \Phi, V, \psi)$ is not a top form {in superspace}, its differential is {in general not} zero. On the contrary, 
the requirement that $\diff {\mathcal L}^{(D|0)}(\Phi, \diff \Phi, V, \psi) =0$ is a strong condition, which {is known to} be achieved in the presence of auxiliary fields. 

To build an action, we have to integrate ${\mathcal L}^{(D|0)}(\Phi, \diff \Phi, V, \psi) $ on the supermanifold and therefore we need to convert it into an integral form  ${\mathcal L}^{(D|N)}$ (for more details see Sec. \ref{Towards full description with integral forms}) as follows:
\begin{eqnarray}
\label{PCOB}
{\mathcal L}^{(D|0)}(\Phi, \diff \Phi, V, \psi)  \longrightarrow  
{\mathcal L}^{(D|N)} = {\mathcal L}^{(D|0)}(\Phi, \diff \Phi, V, \psi) \wedge \mathbb{Y}^{(0|N)}, 
\end{eqnarray}
which is finally integrated on $\mathcal{M}^{(D|N)}$
\begin{eqnarray}
\label{PCOC}
S[\Phi, \diff \Phi, V, \psi] = \int_{\mathcal{M}^{(D|N)}}  {\mathcal L}^{(D|0)}(\Phi, \diff \Phi, V, \psi) \wedge \mathbb{Y}^{(0|N)}. 
\end{eqnarray}
The variational equations obtained from $S[\Phi, \diff \Phi, V, \psi]$ have the 
generic form 
\begin{eqnarray}
\label{PCOD}
\delta S = \int_{\mathcal{M}^{(D|N)}}  \delta \Phi \frac{ \delta {\mathcal L}^{(D|0)} }{ \delta \Phi} \wedge \mathbb{Y}^{(0|N)}. 
\end{eqnarray}
Note that, in deriving \eqref{PCOD}, partial integration is allowed since $\diff  \mathbb{Y}^{(0|N)} =0$ and we 
get the equations 
\begin{eqnarray}
\label{PCODA}
\frac{ \delta {\mathcal L}^{(D|0)} }{ \delta \Phi} \wedge \mathbb{Y}^{(0|N)} =0\,
\end{eqnarray}
on the supermanifold. If $  \mathbb{Y}^{(0|N)} $ has no kernel, we can remove it and obtain the equations of motion 
on the full supermanifold. In general, $  \mathbb{Y}^{(0|N)}$ has a kernel and this implies that there are further solutions to \eqref{PCODA} besides the expected ones. 

The most relevant aspect of the integral \eqref{PCOC} is the reparametrization invariance under all superdiffeomorphisms since it is a top integral form. This translates the powerful technique used in general relativity: using differential forms and integration on top form, one has diffeomorphism invariant quantities. 
In particular, if we consider those superdiffeomorphisms generated by an odd vector $Q$ we can represent the variation as a Lie derivative $ {\mathcal L}_Q $ and we get 
\begin{eqnarray}
\label{PCOE}
0 = \delta_Q S &=&  \int_{\mathcal{M}^{(D|N)}} {\mathcal L}_Q  {\mathcal L}^{(D|0)}  \wedge \mathbb{Y}^{(0|N)} + 
 {\mathcal L}^{(D|0)}  \wedge  {\mathcal L}_Q  \mathbb{Y}^{(0|N)}  \nonumber \\
 &=&  \int_{\mathcal{M}^{(D|N)}}  
 \iota_Q \diff  {\mathcal L}^{(D|0)}  \wedge \mathbb{Y}^{(0|N)} +  {\mathcal L}^{(D|0)}  \diff \iota_Q  \mathbb{Y}^{(0|N)}. 
\end{eqnarray}
Now, three things can happen: 
\begin{enumerate}
\item $\diff {\mathcal L}^{(D|0)}=0$. In this case 
the first term vanishes $\iota_Q \diff  {\mathcal L}^{(D|0)} =0$, but also the second term is zero, by integration by parts. 
It is a common lore, that this can only happen if there are auxiliary fields and using the rheonomic parametrizations satisfying the Bianchi indentities. The latter, however, should not impose the equations of motion, otherwise the action is trivially invariant. 
\item ${\mathcal L}_Q  \mathbb{Y}^{(0|N)} =0$. It means that the Poincar\'e dual $ \mathbb{Y}^{(0|N)}$ is manifestly invariant under supersymmetry and this also implies 
\begin{eqnarray}
\label{PCOF}
 \iota_Q \diff  {\mathcal L}^{(D|0)}  \wedge \mathbb{Y}^{(0|N)} = \diff R^{(D-1|N)}_Q,
\end{eqnarray}
which means that the action is manifestly invariant under supersymmetry, up to a total derivative, in any submanifold described by the Poincar\'e dual $ \mathbb{Y}^{(0|N)}$. This is the powerful construction of superspace actions as in {\cite{gates}}. Since $\mathbb{Y}^{(0|N)}$ is manifestly invariant, the action 
is manifestly invariant. 

\item 
$ \iota_Q \diff  {\mathcal L}^{(D|0)}  \wedge \mathbb{Y}^{(0|N)} = \diff R^{(D-1|N)}_Q$ even in the case that 
${\mathcal L}_Q  \mathbb{Y}^{(0|N)} \neq 0$. This means that, even though $ \mathbb{Y}^{(0|N)} $ is 
not invariant under the supersymmetry, the Lagrangians can be invariant under supersymmetry 
on the bosonic submanifold described by $ \mathbb{Y}^{(0|N)} $. By Eq. \eqref{PCOE} also the last term 
$\int {\mathcal L}^{(D|0)}  \diff \iota_Q  \mathbb{Y}^{(0|N)}$ should vanish.  

\item 
If  $\mathbb{Y}^{(0|N)}$ projects onto the spacetime (see Sec. \ref{Towards full description with integral forms}), then Eq. \eqref{PCOF} implies the supersymmetry on the spacetime. 

\end{enumerate}

Note that Eqs. \eqref{PCOF} do not imply the equations of motion, but 
only that the components of the curvatures along the fermionic directions are expressed 
in terms of the ones along the bosonic directions{, following the principle of rheonomy}.  This is the way in which the invariance of the superspace Lagrangian is realized off-shell. On the other hand, in general (and in the absence of auxiliary fields), the closure of the Bianchi identities, that is the closure of supersymmetry on the fields, also implies the equations of motion, meaning that supersymmetry closes only on-shell. Notice that the same happens on spacetime: the Lagrangian is invariant off-shell, while the supersymmetry algebra closes only on-shell on the fields.

\section{The Geometric Superspace Formulation of the Tensor Multiplet in Rigid $(4,0)$ Theory}\label{Sec3}

The aim of this section is to analyze the main features of a six-dimensional rigid tensor multiplet model on a flat  superspace background, in the chiral theory with USp($4$) $R$-symmetry.
As stated in the Introduction, our construction will be based on the geometric superspace approach \cite{cube}, where all the fields are promoted to form superfields in superspace.

Before introducing the dynamical field content of our theory, let us start by describing the flat {six-dimensional chiral superspace background, which can be found in the low energy limit from a consistent truncation of 11-dimensional supergravity. It is expressed in terms of the following fields 
\begin{equation}
    \left(
    V^a\,,\psi^A\,, B^{[AB]_0}\,, \omega^{ab}\right) \,,
\end{equation}
where $V^a$ ($a=0,1,\ldots,5$) is the vielbein $1$-form, whose bosonic component describes the flat {coordinate frame of the} M5-brane, $\psi^A= -\Gamma_7 \psi^A$ is an antichiral gravitino $1$-form satisfying the pseudo-Majorana condition $\psi^A=\mathbb{C}^{AB} \,C\,\overline\psi_B^t$, with $A=1,\ldots,4 \in$ USp($4$), $B^{[AB]_0}$ are five $2$-form connections [we denote by $[AB]_0$ the irreducible traceless  antisymmetric representation of USp($4$)] and $\omega^{ab}$ is the $\rm{SO}(1,5)$ Lorentz spin connection. They satisfy the following equations defining the background
\begin{eqnarray}
    R^{ab} &\equiv& \diff \omega^{ab}+ \omega^{a}{}_c\, \omega^{cb}=0 \,, \nonumber\\
    T^a &\equiv& DV^a - \frac{\ii}{2} \overline\psi_A \Gamma^a \psi^A=0 \,, \nonumber\\
    \rho^A&\equiv & D\psi^A\equiv \diff \psi^A+ \frac 14 \omega^{ab}\gamma_{ab}\psi^A=0 \,, \nonumber\\
    H^{[AB]_0}&\equiv& \diff B^{[AB]_0} - \ii a_1  \mathbb{C}^{C[A}\overline\psi_C \Gamma_a \psi^{B]_0} V^a =0\,,\label{Bidsbg}
\end{eqnarray}
with $D$ denoting the Lorentz-covariant derivative. An explicit expression of the supervielbein $1$-forms in terms of the coordinates $(x,\theta)$ parametrizing {rigid} superspace, as is well known, is given by
\begin{align}\nonumber
    V^a&=\diff x^a+\frac{\ii}{2}\overline\theta_A\Gamma^a\diff \theta^A,\\
    \psi^A&=\diff \theta^A.
\end{align}
Notice that the fermionic part of the gravitino supervielbein {can be chosen so that it} only has components along the fermionic directions, i.e. $\psi^A_\mu=0$, implying that when spacetime is trivially embedded in superspace, $\theta^A=0$ and $\diff\theta^A=0$, the pullback of the gravitino vanishes. However, there may be more general embeddings in which this does not happen.} \\

The spacetime field content of the six-dimensional  tensor supermultiplet is given by
\begin{equation}
    \left(
    B_{\mu\nu}\,,\lambda_A\,, \phi_{[AB]_0}\right)^I \,,
\end{equation}
where $B^I=\frac{1}{2}\,B^I_{\mu\nu} \diff x^\mu\wedge \diff x^\nu$ ($I=1,\ldots, n$) are $n$ $2$-form connections whose field strengths must satisfy an on-shell anti-self-duality condition on spacetime, $\phi^I_{[AB]_0}\equiv \phi^I_{AB}$ (with $I=1,...,n$) are $5n$ scalars and $\lambda^I_A= +\Gamma_7 \lambda^I_A$ are $n$ chiral  spin-$1/2$ fields, satisfying the pseudo-Majorana condition $\lambda^I_A = -\mathbb{C}_{AB} \,C\,(\overline\lambda^{I B})^t$. Furthermore, $\mu,\,\nu=0,\dots,5$ denote curved spacetime indices. For the complete set of our definitions and conventions see Appendix \ref{conv}.

\subsection{Bianchi identities in superspace and supersymmetry variations of the fields}

The theory under consideration is based on a free differential algebra \cite{cube}, where the  supercurvatures of the dynamical fields are defined  in superspace as follows:
\begin{eqnarray}
    H^I&\equiv&\diff B^I + \ii a_1 \phi^I_{BC} \mathbb{C}^{AC}\overline\psi_A \Gamma_a \psi^B V^a \,, \label{Hdef} \\
    D\lambda^I_A &\equiv  & \diff \lambda^I_A + \frac 14 \omega^{ab}\gamma_{ab}\lambda^I_A \,, \label{Dlambda}\\
    P^I_{AB}&\equiv & \diff \phi^I_{AB} \,.\label{Pphi}
\end{eqnarray}

Imposing the cohomological condition $\diff^2=0$ on the formal definitions \eqref{Hdef}, \eqref{Dlambda}, \eqref{Pphi}, one obtains the following Bianchi identities: 
\begin{align}
    0=&\diff H^I -\ii a_1\,\diff \phi^I_{AB} \mathbb{C}^{AC}\,\overline{\psi}_C\Gamma_a\psi^B\,V^a+2 \ii a_1 \phi^I_{BC} \mathbb{C}^{AC}\overline\psi_A \Gamma_a \rho^B V^a-\ii a_1 \phi^I_{BC} \mathbb{C}^{AC}\overline\psi_A \Gamma_a \psi^B \,{T^a}=\nonumber\\
    =&{\diff H^I -\ii a_1\,\diff \phi^I_{AB} \mathbb{C}^{AC}\,\overline{\psi}_C\Gamma_a\psi^B\,V^a}\,,\label{BI4}\\
   0= &D^2\lambda^I_A - \frac 14 R^{ab}\Gamma_{ab} \lambda^I_A \,, \label{BI5}\\
   0= &D P^I_{AB}= \diff^2 \phi^I_{AB}\,,\label{BI6}
\end{align}
where, in deriving Eq. \eqref{BI4}, we have used the Fierz identity \eqref{4psi} and the expressions \eqref{Bidsbg} for the background fields.
The Bianchi identities \eqref{BI4}-\eqref{BI6} are consistency statements on the formal definitions of the dynamical field strengths. However, they become nontrivial relations among the dynamical degrees of freedom of the theory if we require them to hold in superspace  according to the principle of rheonomy, that is if we endow the field strengths $H^I,D\lambda^I_A, \phi^I_{AB}$ 
with an explicit expansion on a basis of the cotangent bundle of superspace. The latter consists on  requiring the various components along basis elements including odd directions  to be algebraic  functions (in particular, linear tensor combinations) of the ones along entirely bosonic directions. This is what was named \emph{rheonomic parametrization} in \cite{cube}. Besides, the closure of the Bianchi identities \eqref{BI4}-\eqref{BI6} also implies the same equations of motion that will be derived from the Lagrangian in Sec. \ref{Subsec3}. The rheonomic parametrization reads
\begin{eqnarray}
    H^I&= &H^I_{abc}V^aV^bV^c + b_1 \mathbb{C}^{ AB} \overline\psi_A \Gamma_{ab} \lambda^I _BV^aV^b \,, \label{Hpar}\\
    D\lambda^I_A &=  &D_a\lambda^I_A  V^a + b_2 P^I_{AB,a}\Gamma^a \psi^B + b_3 H^I_{abc} \Gamma^{abc} \psi^B \mathbb{C}_{AB} \,, \label{rhpara}\\
    P^I_{AB}&= & P^I_{AB,a}V^a +  \overline\psi_{[A}  \lambda^I _{B]_0}\,, 
\end{eqnarray}
where
\begin{align}
    b_1=\frac{1}{4} a_1\,, \quad b_2=-2\ii\,, \quad b_3=\frac{\ii}{2 a_1}\,.
\end{align}
The value of $a_1$ is fixed by the choice of normalization of the $2$-form $B^I$ and we will choose it to be $a_1=\frac{1}{2}$. The fields   $H^I_{abc}, P^I_{AB,a}$ are usually referred to as the \emph{supercovariant field strengths}.\\

Besides implying the equations of motion, the consistency of the rheonomic parametrizations \eqref{rhpara}
with the Bianchi identities \eqref{BI4}-\eqref{BI6} also requires the anti-self-duality constraint
\begin{equation}
    H^I_{abc}=-\frac{1}{6}\,\epsilon_{abcdef}\,H^{I|def} \,.\label{asd}
\end{equation}
This condition is necessary for the correct on-shell matching of bosonic and fermionic degrees of freedom and in this framework it is not to be imposed by hand, as it follows from the closure of the Bianchi identities in superspace. In particular, it emerges from the sector with two fermionic directions of \eqref{BI4}. This sector yields equations which are equivalent to imposing the closure of supersymmetry transformations on the fields.
It is important to emphasize that this condition does not follow from the spacetime components of the Bianchi identitites alone, once the parametrization of the supercurvatures is chosen. 

From \eqref{Dlambda} and \eqref{rhpara}, one can derive the supersymmetry transformations of the fields, as Lie derivatives along the fermionic directions of superspace
\begin{align}
    \delta_\epsilon B^I&= b_1\,\mathbb{C}^{ AB} \overline\epsilon_A \Gamma_{ab} \lambda^I _BV^aV^b- 2 \ii\,a_1 \phi^I_{BC} \mathbb{C}^{AC}\overline\epsilon_A \Gamma_a \psi^B V^a\,,\nonumber\\
    \delta_\epsilon \lambda^I_A&=b_2 P^I_{AB,a}\Gamma^a \epsilon^B + b_3 H^I_{abc} \Gamma^{abc} \epsilon^B \mathbb{C}_{AB}\,,\label{susytrasfss}\\
    \delta_\epsilon \phi_{AB}^I&=\overline\epsilon_{[A}  \lambda^I _{B]_0}\,,\nonumber
\end{align}
which, on spacetime, after defining
$$\left.H^I_{abc}V^aV^bV^c\right|_{\rm{s.t.}}=\frac{1}{2}\,\partial_\mu B^I_{\nu\rho} \,\diff x^\mu \diff x^\nu\,\diff x^\rho=\frac{1}{3!}\,\mathcal{H}^I_{\mu\nu\rho} \,\diff x^\mu \diff x^\nu\,\diff x^\rho \,,\quad \left.P^I_{AB,a}V^a\right|_{\rm{s.t.}}=\partial_\mu \phi^I_{AB}\diff x^\mu\,,$$
being there $\psi^A_\mu=0$,
reduce to
\begin{align}\label{susytrasfst}
    \delta_\epsilon B^I_{\mu\nu}&= 2 b_1\,\mathbb{C}^{ AB} \overline\epsilon_A \Gamma_{\mu\nu} \lambda^I _B\,,\nonumber\\
    \delta_\epsilon \lambda^I_A&=b_2 \partial_\mu\phi^I_{AB}\Gamma^\mu \epsilon^B + \frac{b_3}2 \partial_\mu B^I_{\nu\rho} \Gamma^{\mu\nu\rho} \epsilon^B \mathbb{C}_{AB}\,,\\
    \delta_\epsilon \phi_{AB}^I&=\overline\epsilon_{[A}  \lambda^I _{B]_0}\,.\nonumber
\end{align}

\subsection{The superspace Lagrangian and its spacetime projection}\label{Subsec3}

The geometric approach allows one to derive the following $(6|0)$-form Lagrangian in superspace, to be integrated on a suitably chosen bosonic submanifold, as previously mentioned. The Lagrangian reads
\begin{align}\nonumber
    \mathcal L^{(6|0)}=&\alpha_1\left(P^I_{AB}-\overline\psi_{[A}\lambda^I_{B]_0}\right)\tilde P_{I \, CD}^aV^{bcdef}\epsilon_{abcdef}\mathbb C^{AC}\mathbb{C}^{BD}-\frac{\alpha_1}{12}\tilde P^I_{AB,l}\tilde P^l_{I \, CD}V^{abcdef}\epsilon_{abcdef}\mathbb C^{AC}\mathbb{C}^{BD}\\ \nonumber
    &+\frac{5\alpha_1}{2}P^I_{AB}\left(\overline\lambda_{I}^A\Gamma_{ab}\psi^B\,V_{cdef}\epsilon^{abcdef}+\frac{4\ii}{5}\phi^I_{CD}\mathbb C^{DA}\overline\psi^B\Gamma_{abc }\psi^{C}V^{abc}\right) \\ \nonumber
    &+ 40 \alpha_1\left(H^I-\frac{1}{8}\overline\psi_A\Gamma_{lm}\lambda^{IA}V^{lm} \right)\tilde H^{abc}_IV^{def}\epsilon_{abcdef}-  \alpha_1\tilde H^I_{lmn}\tilde H^{lmn}_IV^{abcdef}\epsilon_{abcdef}\\ \nonumber
    &- 30\alpha_1 H^I\left(\overline\lambda_{IA}\Gamma_{ab}\psi^AV^{ab}+4\ii\phi_{IAB}\overline\psi^A\Gamma_a\psi^B V^a\right)\\ \nonumber
    &- \frac{\ii \alpha_1}{4}\overline\lambda^I_A\Gamma^a\left( D\lambda_{I}^AV^{bcdef}\epsilon_{abcdef}+\frac{5\ii}{2}\lambda_{BI}\overline\psi^A\Gamma^{bcd}\psi^BV_{abcd}\right)\\ 
    &-\frac{5\alpha_1}{4}\overline\lambda^I_A\Gamma_{abc}\lambda^I_B\overline\psi_{C}\Gamma_d\psi_DV^{abcd}\left(\mathbb C^{AB}\mathbb C^{CD}{+\frac32\mathbb C^{AD}\mathbb C^{BC}}\right)\,, \label{susylag}
\end{align}
where $V^{a_1 a_2 \ldots a_k} \equiv V^{a_1}\wedge V^{a_2} \wedge \dots \wedge V^{a_k}$.
The fields $\tilde P_{I \, AB}^a, \tilde H^I_{abc}$ are auxiliary and will ultimately be identified, through their equations of motion, with the corresponding supercovariant field strengths $ P_{I \, AB}^a,\,H^I_{abc}$ appearing in the superspace parametrizations \eqref{rhpara} of the supercurvatures. They provide a first-order description of the kinetic terms of the corresponding bosonic superfields. This is needed in the present framework, being  our Lagrangian a bosonic $6$-form immersed in the cotangent space of $\mathcal{M}^{(6|16)}$ superspace, in order to avoid the use of the Hodge operator, which is not well defined in this case. 
Note, however, that, using the approach of integral forms, one can define, in a consistent way, the notion of a Hodge-duality operator in superspace.

Moreover, the parameter $\alpha_1$ represents an overall  normalization of the Lagrangian: we fix it as $\alpha_1= -\frac 1{2\cdot 5!}$ in order to have a canonically normalized kinetic term for the scalar fields, when the Lagrangian is projected on spacetime.
 
The spacetime Lagrangian is considerably simpler and reads
\begin{align}\nonumber\label{Lagrangianspacetime} 
    \mathcal L^{{\rm s.t.}}=&\left(\frac 14 \,\partial_\mu\phi^{I AB} \partial^\mu\phi_{I,AB}
  +  \frac{3 }{4}\,\partial_{[\mu} B_{\nu\rho]} \partial^{[\mu} B^{\nu\rho]}
    + \frac{\ii }{8}\overline\lambda^{IA}\Gamma^\mu\, D_\mu\lambda_{IA}\right) \,\diff^6x=\nonumber\\
  =&\left(\frac 14 \,\partial_\mu\phi^{I AB} \partial^\mu\phi_{I,AB}
  +  \frac{1 }{12\, }\,\mathcal{H}^I_{\mu\nu\rho} \mathcal{H}^{I\,\mu\nu\rho}
    + \frac{\ii }{8}\overline\lambda^{IA}\Gamma^\mu\, D_\mu\lambda_{IA}\right) \,\diff^6x\,.
\end{align}

The spacetime Lagrangian is a free Lagrangian for the non interacting fields of the supermultiplet and it is invariant under the supersymmetry transformation in \eqref{susytrasfst} up to a total derivative:
\begin{equation}
    \delta_\epsilon \mathcal{L}^{{\rm s.t.}}= \partial_\mu K^\mu \,\diff^6x \,,\label{deltalst}
\end{equation}
with 
\begin{equation}
  K^\mu=\frac14\left(\overline\lambda^{IA}\Gamma^{\nu}\Gamma^\mu\epsilon^B\partial_\nu\phi_{IAB}-\frac14\overline\lambda^{I}_A\Gamma^{\rho\sigma\tau}\Gamma^{\mu}\epsilon^A\partial_\rho B_{I\sigma\tau}\right)\,.
  \label{cur}
\end{equation}
Off-shell invariance of the spacetime Lagrangian under supersymmetry implies the presence of a conserved Noether current, which reads 
\begin{equation}
    \mathcal{J}^\nu_A=-\frac12\Gamma^\mu\Gamma^\nu\lambda^{IB}\partial_\mu\phi_{ABI} +\frac18 \Gamma^{\rho\sigma\tau}\Gamma^{\nu}\lambda^I_A\partial_\rho B_{\sigma\tau I} \,. 
\end{equation}
One can see that it is indeed conserved $\partial_\nu\mathcal{J}^\nu_A=0$, upon the use of the equations of motion.
Notice that this invariance property does not require the anti-self-duality condition on the tensor field strengths, which is, however, necessary for closure of supersymmetry on the fields and thus on the Lagrangian. {Indeed, the spacetime Lagrangian \eqref{Lagrangianspacetime} depends on both the self-dual and the anti-self-dual parts of $\partial_{[\mu}B^I_{\nu\rho]}$, but the self-dual component only enters the supersymmetry variation of \eqref{Lagrangianspacetime} in the total derivative term \eqref{deltalst}.} {However, the equations of motion involve both the self-dual and the anti-self-dual parts of $\partial_{[\mu}B^I_{\nu\rho]}$, thus leading to unmatched propagating degrees of freedom.}

We emphasize that, in the geometric approach pursued in the present paper, the anti-self-duality condition is not imposed by hand, but follows from the closure of the Bianchi identities, {and also, independently, from the Euler-Lagrange} equations {in superspace} derived from the  superspace Lagrangian $6$-form \eqref{susylag}, along the fermionic directions of superspace. 
This is an instance of the general property that the Euler-Lagrange equations in superspace encode far more information than their restriction to spacetime.\\
Let us conclude this section by listing the Euler-Lagrange equations  coming from the Lagrangian \eqref{susylag}, which are tensorial form equations in superspace, with a short account of their implications in both even and odd directions.   \\
The components along the bosonic vielbein $V^a$ give the standard field equations of the dynamical fields on spacetime, whereas the components along directions including at least one odd vielbein $\psi^A$ are constraints, some of which are Fierz identities among the spinorial fields, that are identically satisfied, while the rest are constraints on the field strengths of the dynamical fields  that have to be satisfied on-shell, among which the anti-self-duality condition \eqref{asd} on the supercovariant field strength of the $2$-form potential. {The constraints resulting from the Bianchi identities are in agreement with the Euler-Lagrange equations in superspace.}

\subsubsection{The equations of motion of the auxiliary fields}
The equations of motion for $\tilde H^I_{abc}$ and $\tilde P^I_{AB,a}$ imply the following identifications:
\begin{align}
    \tilde H^I_{abc}=H^I_{abc}\,,\qquad \tilde P^I_{AB,a}=P^I_{AB,a}\,.
\end{align}

\subsubsection{Equations of motion of $B^I$}

The equations of motion for the field $B^I$ are
\begin{align}
    &-40 \, \diff \tilde H^{abc}_IV^{def}\epsilon_{abcdef}-60\ii\tilde H^{abc}_I\overline\psi_A\Gamma^d\psi^AV^{ef}\epsilon_{abcdef}+ 30(\overline{\mathcal D\lambda}_{IA}\Gamma_{ab}\psi_B)V^{ab}\mathbb C^{AB}\\ \nonumber
    &-30 \ii(\overline\lambda_{IA}\Gamma_{ab}\psi_B)(\overline\psi_C\Gamma^a\psi^C)V^b\mathbb C^{AB}+120 \ii\diff \phi_{IAB}(\overline\psi^A\Gamma_a\psi^B)V^a-60\phi_{IAB}(\overline\psi^A\Gamma_a\psi^B)(\overline\psi_C\Gamma^a\psi^C)=0 \,.
\end{align}
\begin{itemize}
    \item The sector $V^4$ gives
    \begin{align}
            \partial^a \tilde H^I_{abc}=0\quad \Rightarrow
            \quad \partial^a H^I_{abc}=0\,.
    \end{align}
    {The above equation, which matches what one would obtain from the Bianchi identities, describes the dynamics of $B^{I}$.}
    \item The sector $\psi V^3$ relates the spinorial derivative of the supercovariant field strength with the spacetime derivative of the spinor field
    \begin{align}
        &\nabla_B\tilde H^I_{abc}=-\frac 18\Gamma_{[ab}\mathcal D_{c]}\lambda^I_B \,,
    \end{align}
    where the spinorial derivative $\nabla_A$ is defined as $\diff=V^a\partial_a+\overline\psi^A\nabla_A$. This result once again coincides with the one coming from the Bianchi identities once we impose the anti-self-duality condition \eqref{asd}.
    \item The sector $\psi^2 V^2$ gives a relation between terms containing $H^I_{abc}$ which is satisfied only if the anti-self-duality condition on $H^I_{abc}$
    \begin{equation}
    H^I_{abc}=-\frac{1}{6}\,\epsilon_{abcdef}\,H^{I|def} \,\label{asd'}
\end{equation}
    holds.
    
    \item The sector $\psi^3 V$ is automatically satisfied due to Fierz identities among the spinors.
    \item The sector $\psi^4$ leads to 
    \begin{align}
        \mathbb \phi_{IAB}(\overline\psi^A\Gamma_a\psi^B)(\overline\psi_C\Gamma^a\psi^C)=0 \,,
    \end{align}
    which vanishes thanks to the Fierz identity \eqref{4psi}. 
\end{itemize}

\subsubsection{Equations of motion of $\overline{\lambda}_{IA}$}
\begin{align}\nonumber
    &-\psi_B P^a_{CD \, I}V^{bcdef}\epsilon_{abcdef}\mathbb C^{AC}\mathbb C^{BD} +5 \mathbb C^{AB}\Gamma_{lm}\psi_BH^{abc}_IV^{lmdef}\epsilon_{abcdef}\\ \nonumber
    &- 30 \,H_I\Gamma_{ab}\psi^AV^{ab}-\frac\ii 2\,\Gamma^a\mathcal D\lambda_I^A V^{bcdef}\epsilon_{abcdef}+\frac{5}{8}\Gamma^a\lambda^A_I(\overline\psi_E\Gamma^b\psi^E)V^{cdef}\epsilon_{abcdef}\\ \nonumber
    &-\frac52\,\mathbb C^{EA}\Gamma_{ab}\psi^F P_{I \, EF}V_{cdef}\epsilon^{abcdef}\\
     &{+\Gamma_{abc}\lambda_{IB}(\overline\psi_{C}\Gamma_d\psi_D)V^{abcd} \left[-\frac 52\mathbb C^{AB}\mathbb C^{CD}{+\frac{15}8\mathbb C^{AC}\mathbb C^{BD}-\frac{15}8\mathbb C^{AD}\mathbb C^{BC}} \right]}\nonumber\\
    &{+\frac 54\Gamma_a\lambda_{BI}(\overline\psi_C\Gamma_{bcd}\psi_D)V^{abcd}\mathbb C^{AC}\mathbb C^{BD}} =0 \,.
\end{align}

\begin{itemize}
    \item The sector $V^6$ leads to the equations of motion for the spin-$1/2$ field $\lambda_I^A$ as expected
    \begin{align}
        \slashed{\mathcal D}\lambda^A_I=0 \,.
    \end{align}
    \item The sector $\psi V^5$ again leads to an identity that can only be satisfied if $H^I_{abc}$ is anti-self-dual.
    \item The sector $\psi^2 V^4$ is identically satisfied with the given coefficients, due to Fierz identities.
\end{itemize}

\subsubsection{Equations of motion of $\phi^I_{AB}$}
\begin{align}\nonumber
    &-\diff \tilde P^a_{CD \, I} V^{bcdef}\epsilon_{abcdef}\mathbb C^{AC}\mathbb C^{BD}-\frac{5\ii}{2}\tilde P^a_{CD \, I}(\overline\psi_E\Gamma^b\psi^E)V^{cdef}\epsilon_{abcdef}\mathbb C^{AC}\mathbb C^{BD}\\ \nonumber
    &+20 \ii (\overline\psi^{[A}\Gamma_l\psi^{B]_0})H^{abc}_IV^{ldef}\epsilon_{abcdef}+15 (\overline\psi^{[A}\Gamma_l\psi^{B]_0})(\overline\lambda_{EI}\Gamma_{ab}\psi_F)V^{abl}\mathbb C^{EF}\\ \nonumber
    &{{+} 60\ii  (\overline\psi^{[A}\Gamma_l\psi^{B]_0})\phi_{IEF}(\overline\psi^E\Gamma_a\psi^F)V^{la}}-120 \ii\,H^I(\overline\psi^{[A}\Gamma_a\psi^{B]_0})V^a\\
    &-\frac 52\,(\mathcal D\overline\lambda^{[A}_I\Gamma_{ab}\psi^{B]_0})V_{cdef}\epsilon^{abcdef}+5\ii\,(\overline\lambda^{[A}_I\Gamma_{ab}\psi^{B]_0})(\overline\psi_E\Gamma_c\psi^E)V_{def}\epsilon^{abcdef} \nonumber \\
     &{+40\ii\, P_{I \, DC} \left( \overline \psi^{[A} \Gamma_{abc} \psi^{|C|} \right) V^{abc} \mathbb C^{B]_0 D} {+} 30\,\phi_{I \, DC} \left( \overline \psi^D \Gamma_{abc} \psi^{[B} \right) \left( \overline \psi_G \Gamma^a \psi^{|G|} \right) V^{bc} \mathbb C^{|C|A]_0}}  = 0 \,.
\end{align}
\begin{itemize}
    \item The sector $V^6$ leads to the Klein-Gordon equation for the scalar field
    \begin{align}
        \Box\phi^I_{AB}=0 \,.
    \end{align}
    \item The sector $\psi V^5$ yields the following relation between the spinorial derivative of $P^I_{CD,a}$ and the spacetime derivative of $\lambda^I_A$
    \begin{align}
        \nabla^C P^I_{AB,a}=-\delta^C_{[A}\mathcal D_a\lambda^I_{B]_0} \,.
    \end{align}
    \item The sector $\psi^2 V^4$, as it happened for the other equations of motion, can only be satisfied if \eqref{asd} holds.
    \item The sectors $\psi^3V^3$ and $\psi^4V^2$ are satisfied thanks to Fierz identities.
\end{itemize}

\section{Retrieving Sen's Lagrangian and its Superspace Extension}\label{Sec4}

{The main goal of the present} investigation is the construction of an M5-brane, noninteracting Lagrangian in superspace which would yield, when restricted to spacetime, the description given by Sen of the same physical system (see also \cite{Lambert:2019diy}). In fact this theory is chosen as a {simplified model} in order to devise a more general prescription for achieving an extension of Sen's description of chiral forms to superspace.
The aim of this section is therefore to modify the superspace Lagrangian $6$-form \eqref{susylag} so that:
\begin{enumerate}
     \item Once restricted to spacetime, it yields Sen's description of the same system (or an equivalent version of it);
     \item Its Euler-Lagrange equations in superspace give the superspace constraints (rheonomic) for the physical fields, besides yielding the supercovariant equations of motion in spacetime.
\end{enumerate}
To attain points 1 and 2 above, it is useful to rewrite the Lagrangian $6$-form in \eqref{susylag} in the following, more compact, way:
\begin{equation}\label{rheold}
    \mathcal{L}^{{(6|0)}}=(\diff B^I+Z^I)\wedge{}^*\tilde{H}_I-\frac{1}{2}\,\tilde{H}^I\wedge{}^*\tilde{H}_I+\diff B^I\wedge Z_I+ \mathcal{L}^{{(6|0)}}_i(\Phi)\,,
\end{equation}
where we have generically denoted by $\Phi$ the scalar and spin-$1/2$ fields, so that  $ \mathcal{L}^{{(6|0)}}_i(\Phi)$ does not depend either on the $2$-form or on $\tilde{H}_{I\,abc}$. Moreover  we have defined:
\begin{align}
  \tilde{H}^I&\equiv  \tilde{H}^I_{abc}\,V^a\wedge V^b\wedge V^c\,,\nonumber\\
  Z^I&=Z^I(\Phi)\equiv  \frac{1}{8}\, \overline{\lambda}_A^I \Gamma_{ab} \psi^A\,V^a\wedge V^b + \frac{\ii}{2} \phi^I_{AB}\,\overline\psi^A \Gamma_a \psi^B V^a \,.\label{ZI}
\end{align}
From Eqs. \eqref{Hdef} and \eqref{Hpar}, we find {(when the Bianchi identities in superspace are satisfied)}:
\begin{equation}
    \diff B_I+Z_I=H_I\equiv H_{I\,abc}\,V^a\wedge V^b\wedge V^c\,,\label{Bpara2}
\end{equation}
and the Bianchi identities in superspace imply the anti-self-duality \eqref{asd} of $H^I_{abc}$. It is straightforward to verify that the {Euler-Lagrange} equations for $\tilde{H}_I$ and for $B^I$ read:
\begin{align}
  \tilde{H}_I&=  \diff B_I+Z_I=H_I\,,\label{tilde}\\
  0&=\diff\left({}^* \tilde{H}+Z\right)\,.
\end{align}
The last equation is satisfied using the first one, Eq. \eqref{Bpara2}, and the anti-self-duality of $H^I$. The variation of $\mathcal{L}^{{(6|0)}}$ with respect to the other fields $\Phi$ yields:
\begin{equation}
    \delta_{{\Phi}} \mathcal{L}^{{(6|0)}}=-\delta_{{\Phi}} Z_I\wedge \left[2\,H^I-Z^I\right]+\delta_{{\Phi}}\mathcal{L}^{{(6|0)}}_i\,,\label{varest}
\end{equation}
where $\delta_{{\Phi}} Z_I=\frac{\delta Z_I}{\delta \Phi}\,\delta \Phi$. Our theory is noninteracting since $Z_I$, having only components along $\psi V$ and $\psi\psi$, vanishes when restricted to spacetime ($\theta=0=\diff \theta$). \par
Before setting out to extend Sen's prescription to superspace in order to formulate a Lagrangian 6-superform satisfying the above points 1 and 2, we wish to first review the construction by Sen in a specific class of bosonic theories describing chiral forms on spacetime, and suggest an equivalent first-order formulation which will be instrumental for our purposes. The reason for this, which we anticipate here, is that Sen's prescription requires the introduction of new fields $P^I$ which appear in the Lagrangian in terms of the form $\diff P^I\wedge {}^* \diff P_I$. A straight superspace extension of these terms requires a consistent definition of the Hodge operator ${}^*$ in superspace, which was achieved within the framework of integral forms \cite{Castellani:2014goa,Castellani:2015paa,Castellani:2016ibp}. This formulation of the problem will be discussed in the last section. In the present section we wish to follow a different route. The definition of a Hodge
duality operator, which seems to be necessary in order to write the kinetic term of the bosonic
fields $P^I$, can be eluded by introducing a $0$-form tensor field as is usual in the first-order approach to the kinetic terms. This is indeed what we did in writing the kinetic terms of
the $2$-form $B_I$ and of the scalar fields $\phi_{[AB]}^I$ in the Lagrangian of the (noninteracting) M5-brane in Sec. \ref{Sec3}.
It follows that a possible way of extending Sen's construction to superspace is to change the corresponding Lagrangian
into a completely equivalent  one, albeit the duality operator is replaced by a first-order formulation.

\subsection{Review of Sen's construction and its first-order formulation}
Let us review Sen's prescription for a particular bosonic theory in a $(4n+2)$-dimensional spacetime, describing chiral $(2n)$-forms $B_I$ whose field strengths
\begin{equation}
    H_I\equiv \diff B_I+Y_I\,,
\end{equation}
are required to be anti-self-dual,
$$H_I=-{}^* H_I\,.$$
An example of a model of this kind is that of Type IIB theory in which the metric is frozen to be flat and the fermionic fields are set to zero, which is discussed in the first part of \cite{Sen:2015nph}. In that case, $n=2$ and there is just one chiral $4$-form $B$ and $Y\equiv B^{(2)}\wedge F^{(3)}$. As opposed to the type IIB example {discussed in \cite{Sen:2015nph}}, here we require the  corresponding field strengths to be anti-self-dual instead of self-dual.\par 
Let us now consider the following class of $(4n+2)$-form \textit{spacetime} Lagrangians:
\begin{equation}
   \mathcal{L}=(\diff B^I+Y^I)\wedge{}^*\tilde{H}_I-\frac{1}{2}\,\tilde{H}^I\wedge{}^*\tilde{H}_I+\diff B^I\wedge Y_I+ \mathcal{L}_i(\Phi)\,,  \label{Lst0}
\end{equation}
which includes our bosonic model for $n=1$. Furthermore, note the formal analogy between the above Lagrangian and the one in \eqref{rheold}. The difference, however, is that the Lagrangian in \eqref{rheold} is a $6$-form in \textit{superspace} and $Z_I$ are superspace-$3$-forms with vanishing spacetime restriction. Nevertheless this formal analogy will guide us in the next section in formulating a superspace Lagrangian for our supersymmetric model meeting the requirements 1 and 2 above.\par
Applied to a Lagrangian of the form \eqref{Lst0}, Sen's prescription would yield
\begin{equation}
   \tilde{\mathcal{L}}=-\left[\frac{1}{2}\,\diff P^I \wedge{}^*\diff P_I+ (\diff P^I+Y^I)\wedge Q_{I-}-\frac{1}{2}\,Y^I\wedge {}^* Y_I \right]+ \mathcal{L}^{{\rm st}}_i(\Phi)\,, \label{Lst1}
\end{equation}
where
\begin{equation}
    Q^I_-\equiv Q^I_{-abc}\,V^a\wedge V^b\wedge V^c=-{}^* Q^I_-\, 
\end{equation}
is an auxiliary anti-self-dual {$(2n+1)$}-form, and $P^I$ new 2{$n$}-forms. {Let us mention that, in this subsection, since we consider a purely bosonic theory in flat spacetime, we work with the bosonic vielbein $V^a=\diff x^a$.} Note that the kinetic terms for the $P^I$ fields have the wrong sign.\footnote{Recall that we are using the "mostly minus" convention and $\frac{1}{2}\,\omega_{(3)} \wedge{}^* \omega_{(3)}=\frac{1}{2\,3!}\,\omega_{abc}\,\omega^{abc}\,\diff^6 x$, where $\omega^{(3)}\equiv \frac{1}{3!}\,\omega_{abc}\,V^a\wedge V^b\,\wedge V^c$ and $\diff^6 x\equiv- \frac{1}{6!}\,V^{a_1}\wedge\dots V^{a_6}\,\epsilon_{a_1\dots a_6}$.}
The field equations read \cite{Sen:2015nph}
\begin{align}
    \frac{\delta \tilde{\mathcal{L}}}{\delta Q^I_-}=0&\Leftrightarrow\,\,\,\mathbb{P}_+(\diff P^I+Y^I)=0\,,\label{PpdpY}\\
       \frac{\delta \tilde{\mathcal{L}}}{\delta P^I}=0&\Leftrightarrow\,\,\,\diff\left({}^* \diff P^I+Q_-^I\right)=0\,,\label{dsPQ}
\end{align}
where {we denote by} $\mathbb{P}_\pm$ the projectors to the self- and anti-self-dual components of a {$(2n+1)$}-form, respectively.
Equations \eqref{dsPQ} are solved by equating ${}^* \diff P^I+Q_-^I$ to an exact form. It is useful to choose the latter in the following two equivalent ways:
\begin{align}
    -\diff P^I+{}^* \diff P^I+Q_-^I&=2\,\diff \Xi_1^I\,,\label{S1}\\
    \diff P^I+{}^* \diff P^I+Q_-^I&=2\,\diff \Xi_2^I\,.\label{S2}
\end{align}
where we have introduced two sets of forms $\Xi^I_1,\,\Xi^I_2$ related as follows: $\Xi_1^I=\Xi_2^I-P^I$. From Eq. \eqref{S1} it follows that
$$\mathbb{P}_+(\diff \Xi_1^I)=0\,\Rightarrow\,\,\,\diff \Xi_1^I=-{}^* \diff \Xi_1^I\,\Rightarrow\,\,\,\diff {}^* \diff \Xi_1^I=0\,,$$
namely the forms $\Xi^I_1$ are free and decouple from all the other fields. The forms $\Xi^I_2$, on the other hand, are interacting and can be identified with the physical forms $B^I$. Indeed from Eqs. \eqref{S2} and \eqref{PpdpY} we find
\begin{equation}
 \mathbb{P}_+(\diff \Xi_2^I)=  \mathbb{P}_+(\diff P^I)=-\mathbb{P}_+(Y^I)\,\,\Rightarrow\,\,\,\mathbb{P}_+(\diff \Xi_2^I+Y^I)=0\,, \label{p+2}
\end{equation}
and 
\begin{equation}
    \mathbb{P}_-(\diff \Xi_2^I+Y^I)=\frac{Q^{{I}}_-}{2}+\mathbb{P}_-(Y^I)\,.
\end{equation}
Upon identifying $B^I= \Xi_2^I$ and the corresponding field strength $H^I$ as
\begin{equation}
    H^I\equiv \diff B^I+Y^I\,,
\end{equation}
and using \eqref{p+2}, we find
\begin{equation}
   2H^I= 2\,\mathbb{P}_-(H^I)=Q^I_-+2\,\mathbb{P}_-(Y^I)\,,\label{2HQ}
\end{equation}
which corresponds to Eq. (3.16) of \cite{Sen:2015nph}. Then, computing the variation of $\tilde{\mathcal{L}}$ with respect to the other fields $\Phi$, one finds
\begin{equation}
  \delta_{{\Phi}} \tilde{\mathcal{L}}=-\delta_{{\Phi}} Y_I\wedge \left[Q_-^I-{}^* Y^I\right]+\delta_{{\Phi}}\mathcal{L}_i{^{{\rm st}}}=-\delta_{{\Phi}} Y_I\wedge \left[2\,H^I- Y^I\right]+\delta_{{\Phi}} \mathcal{L}_i{^{{\rm st}}}\,,
\end{equation}
which coincides with the corresponding variation of the Lagrangian  ${\mathcal{L}}$ {in \eqref{Lst0}, once one passes to second order for $\tilde H^I$, expressing it in terms of $H^I$ [see the analogous Eq. \eqref{tilde}] which, as shown in \eqref{2HQ}, is anti-self-dual}.\par
As discussed above, here we wish to rewrite the kinetic terms for the $P_I$ fields in \eqref{Lst1} in an equivalent first-order form which will be instrumental to the application, in the next subsection, of an appropriate extension of Sen's construction to the superspace Lagrangian \eqref{rheold}. To this end, we 
introduce the following auxiliary fields:
\begin{equation}
    \tilde{H}^{{I}}\equiv \tilde{H}^{{I}}_{abc}\,V^a\wedge V^b\wedge V^c \,,\,\,\,
  \hat{H}^I_-= \hat{H}^I_{-\,abc}\,V^a\wedge V^b\wedge V^c=-{}^* \hat{H}^I_- \,
\end{equation}
and write the following Lagrangian $(4n+2)$-form in spacetime {which, as we are going to show in the following, is the first-order formulation of Sen's Lagrangian \eqref{Lst1}:}
\begin{equation}
   \tilde{\mathcal{L}}'=-\left[(\diff P_I+Y_I)\wedge \tilde{H}^{{I}}+\tilde{H}^{{I}}\wedge \hat{H}_{I-}+Y_I\wedge \hat{H}^I_{-} \right]+ \mathcal{L}^{{\rm st}}_i(\Phi)\,. \label{Lst2}
\end{equation}
We wish to prove that $\tilde{\mathcal{L}}'$ is equivalent to $\tilde{\mathcal{L}}$. To this end we compute the field equations from {$\tilde{\mathcal{L}}'$}, which read
\begin{align}
     \frac{\delta  \tilde{\mathcal{L}}'}{\delta \tilde{H}^I}=0&\Leftrightarrow\,\,\,\diff P^I+Y^I=\hat{H}^I_-\,,\label{PpdpY2}\\
       \frac{\delta  \tilde{\mathcal{L}}'}{\delta \hat{H}^I_-}=0&\Leftrightarrow\,\,\,
    \mathbb{P}_+(\tilde{H}^I)= - \mathbb{P}_+(Y^I)\,,\label{cHY}\\
       \frac{\delta  \tilde{\mathcal{L}}'}{\delta P^I}=0&\Leftrightarrow\,\,\,\diff\left(\tilde{H}^I\right)=0\,.\label{dQmQp}   
\end{align}
Equation \eqref{PpdpY2} clearly implies that
\begin{equation}\label{py}
    \mathbb{P}_+(\diff P^I+Y^I)=0\,\,,\,\,\,\mathbb{P}_-(\diff P^I+Y^I)=\hat{H}^I_-\,.
\end{equation}
Equation \eqref{dQmQp} is, as usual, solved by equating $\tilde{H}^I$ to exact forms, namely by introducing  a new set of forms $\Xi^I$ and  setting
\begin{equation}\label{tildeh}
 \tilde{H}^I=\diff \Xi^I\,\,\leftrightarrow\,\,\, \mathbb{P}_-(\tilde{H}^I)=-\mathbb{P}_+(\tilde{H}^I)  +\diff \Xi^I=\mathbb{P}_+(Y^I) +\diff \Xi^I\,,
\end{equation}
where we have used \eqref{cHY}. From the above relations we find
\begin{align}
    \mathbb{P}_-(\diff \Xi^I)= \mathbb{P}_-(\tilde{H}^I)\,,\,\,\,\, \mathbb{P}_+(\diff \Xi^I+Y^I)=0\,.\label{PpdXiY}
\end{align}
We now define the following sets of fields:
\begin{equation}
    B^I\equiv \frac{P^I+\Xi^I}{2}\,\,,\,\,\,\tilde{P}=\frac{P^I-\Xi^I}{2}\,,
\end{equation}
where $B^I$ are the physical forms.
From the second of Eqs. \eqref{PpdXiY} and  the first of Eqs.  \eqref{py} it follows that
\begin{equation}
      H^I\equiv \diff B^I+Y^I=\mathbb{P}_-( H^I)\,,\,\,\,\,\mathbb{P}_+(\diff\tilde{P}^I)=0\,.
\end{equation}
From this we conclude that the {$2n$}-forms $\tilde{P}^I$ are free. They indeed coincide with the fields $-\Xi_1^I=P^I-B^I$ introduced earlier {in Eq. \eqref{S1}}. {Equation} \eqref{PpdpY2}{, the second of Eqs. \eqref{py}}  and the first of Eqs. \eqref{PpdXiY}, on the other hand, allow us to write
\begin{equation}
    2\,H^I= \hat{H}^I_-+\mathbb{P}_-(\tilde{H}^I+Y^I)\,.\label{HhH}
\end{equation}
Comparing the above equation with \eqref{2HQ} we derive the following relation  {between the  auxiliary fields of the original second-order Lagrangian description and the ones in the  present first-order formulation}:
\begin{equation}
    Q_-^I=\hat{H}^I_-+\mathbb{P}_-(\tilde{H}^I-Y^I)\,.\label{QH}
\end{equation}
 Finally let us compute the variation of the Lagrangian with respect to $\Phi$:
\begin{equation}
  \delta_\Phi \tilde{\mathcal{L}}'=-\delta_\Phi Y_I\wedge \left[\hat{H}^I_-+\tilde{H}^I\right]+\delta_\Phi\mathcal{L}_i=-\delta_\Phi Y_I\wedge \left[2\,H^I- Y^{{I}}\right]+\delta_\Phi\mathcal{L}_i{^{\rm st}}\,,
\end{equation}
where we have used \eqref{HhH} and \eqref{cHY}. We see that $\delta_\Phi \tilde{\mathcal{L}}'=\delta_\Phi \tilde{\mathcal{L}}$ once the auxiliary fields are expressed in terms of the dynamical ones.\par
Let us now comment on the off-shell equivalence between $\tilde{\mathcal{L}}$ and $\tilde{\mathcal{L}}'$. 
The first-order formulation of  $\tilde{\mathcal{L}}$  is effected by introducing two new sets  of auxiliary fields $\hat{H}^I_-$ and $\mathbb{P}_+(\tilde{H}^{{I}})$. The equation of the former is \eqref{cHY}, while the equation of the latter is 
\begin{equation}\mathbb{P}_-(\diff P^I+Y^I)=\hat{H}^I_-\,.\label{eqHtp}
\end{equation}
Eliminating these extra auxiliary fields using their equations of motion \eqref{cHY} and \eqref{eqHtp}, and relating $\mathbb{P}_-(\tilde{H}^I)$ to $Q_-^I$ through \eqref{QH},
\begin{equation}
    Q_-^I=\hat{H}^I_-+\mathbb{P}_-(\tilde{H}^I-Y^I)=\mathbb{P}_-(\tilde{H}^I+\diff P^I)\,,\label{QH2}
\end{equation}
the reader can derive $\tilde{\mathcal{L}}$ from $\tilde{\mathcal{L}}'$.\footnote{Note that the relation between Sen's auxiliary field $Q^I_-$ and the one coming from our first-order formulation, $\tilde H^I_-$, is cohomologically nontrivial, since their difference is not exact.}
We therefore conclude that the Lagrangians $\tilde{\mathcal{L}}$ and $\tilde{\mathcal{L}}'$ are equivalent.

\subsection{Extending Sen's construction to superspace}
In this section, we shall use the general first-order expression of \eqref{Lst2} as inspiration in order to devise a Lagrangian 6-superform $\hat{\mathcal{L}}$, equivalent to the superspace Lagrangian \eqref{susylag}, describing the noninteracting M5-brane and satisfying point 1 and 2 outlined earlier.

Let us first give some definitions. Writing a generic $3$-form in superspace as  $$\Omega=\Omega^{(3,0)}+\Omega^{(2,1)}+\Omega^{(1,2)}+\Omega^{(0,3)}\,,$$
where the four terms on the right-hand side are the components of $\Omega$ along $VVV,\,VV\psi,\,V\psi\psi,\psi\psi\psi$, respectively, let us define the action of operators $\mathbb{P}^{abc}_\pm$ on a 3-superform $\Omega$ in superspace as the projections of the only $(3,0)$ component of $\Omega$ into its self- and anti-self-dual components, respectively, leaving all other superspace components of $\Omega$ unaltered. The equation
\begin{equation}
    \mathbb{P}^{abc}_\pm ( \Omega)=0\,\,\Leftrightarrow\,\,\,\,\left(V^a\wedge V^b\wedge  V^c\mp\frac16\, \epsilon^{defabc} (V_d\wedge V_e\wedge V_f)\right)\wedge \Omega=0\,
\end{equation}
therefore implies that the self- or anti-self-dual part of $\Omega^{(3,0)}$, respectively, vanish, while the other superspace components of the same form must vanish separately:
\begin{align}
    \mathbb{P}^{abc}_\pm ( \Omega)=0  \,\,\Leftrightarrow\,\,\,\,\begin{cases}\mathbb{P}_\pm(\Omega^{(3,0)})\equiv \frac{1}{2}\,\left(\Omega_{abc}\pm \frac{1}{6}\,\epsilon_{abcefg}\Omega^{efg}\right)\,V^a\wedge V^b\wedge V^c=0 \,, \cr 
    \Omega^{(1,2)}=\Omega^{(2,1)}=\Omega^{(0,3)}=0\,.\end{cases}\,\,\label{defPP}
\end{align}
We now introduce the following set of  auxiliary fields in superspace:
\begin{align}
    \tilde{{\bf H}}^I&=\tilde{H}^I+\Delta \tilde{H}^I\,\,,\,\,\,\hat{H}^I_{-}{}_{abc}=-\frac16\epsilon_{abcdef }\hat{H}^{Idef}_-\,,
\end{align}
Let us define, for notational convenience, $\hat{H}_-^I\equiv \hat{H}^I_{-\,abc}\,V^a\wedge V^b\wedge V^c$, so that $\hat{H}^I_-=-{}^*\hat{H}^I_-$, i.e. $\mathbb{P}_+(\hat{H}^I_-)=0$, $\mathbb{P}_\pm$ being defined on $(3,0)$ components of $3$-forms as in \eqref{defPP}.\par
Differently from the previously described spacetime description, $\tilde{{\bf H}}^I$ is now a superfield with $(3,0)$ components $\tilde{{ H}}^I$, and $(1,2)$, $(2,1)$, and $(0,3)$ components encoded in $\Delta \tilde{{ H}}^I$.\par
Let us write the Lagrangian $6$-form in superspace of the same general expression \eqref{Lst2}, namely as follows:
\begin{equation}\label{eqL2A}
    \hat{\mathcal{L}}=-\left((\diff P_I+Z_I)\wedge  \tilde{{\bf H}}^I+\tilde{{\bf H}}_I\wedge \hat{H}^I_- +Z_I\wedge \hat{H}_-^I\right)+ \mathcal{L}_i(\Phi)\,,
\end{equation}
where $Z_I=Z_I(\Phi)$ are given by \eqref{ZI}.\par
Let us now compute the field equations in superspace:
\begin{align}
    \frac{\delta \hat{{\mathcal{L}}}}{\delta \tilde{{\bf H}}^{{I}}}=0\,&\Leftrightarrow
    \,\, \diff P^I+Z^I=\hat{{ H}}^I_-\,,\label{eeq2}\\
   \frac{\delta\hat{{\mathcal{L}}}}{\delta \hat{H}^I_{-\,abc}}=0\,&\Leftrightarrow\,\,   \mathbb{P}^{abc}_+(\tilde{{\bf H}}^I+Z^I)=0\,,\label{eeq5}\\
   \frac{\delta\hat{{\mathcal{L}}}}{\delta P^I}=0\,&\Leftrightarrow\,\, \diff \tilde{{\bf H}}^I=0\,.\label{eeq4}
\end{align}
Equation \eqref{eeq2} implies that $\diff P^I+Z^I$, being equal to $\hat{H}^I_-$, is a $(3,0)$-form. It therefore makes sense to compute on them the projectors $\mathbb{P}_\pm$, defined in \eqref{defPP}, so that we have:
\begin{equation}
     \mathbb{P}_+(\diff P^I+Z^I)=0 \,,\,\,\,\mathbb{P}_-( \diff P^I+Z^I)=\mathbb P_-(\hat{{ H}}^{I}_-)\,.\label{eeq1}
\end{equation}
We solve equation \eqref{eeq4} by equating $ \tilde{{\bf H}}^I$ to exact forms in superspace:
\begin{equation}
\tilde{{\bf H}}^I=\diff \Xi^I\,\,\Leftrightarrow\,\,\, \mathbb{P}_-(\tilde{H}^I)=-\mathbb{P}_+(\tilde{H}^I)-\Delta \tilde{H}^I+\diff \Xi^I \,.\label{eeq42}
\end{equation}
Equation \eqref{eeq5} implies
\begin{equation}
    \mathbb{P}_+(\tilde{{ H}}^I)=0\,\,,\,\,\,\,\Delta \tilde{H}^I=- Z^I\,,\label{eeq52}
\end{equation}
since $Z^I$ have vanishing $(3,0)$ components.
From this and applying  $\mathbb{P}_+^{abc}$ to both sides of Eq. \eqref{eeq42}, we find
\begin{equation}
     \mathbb{P}_+^{abc}\left(\diff \Xi^I- \tilde{{\bf H}}^I\right)=0\,.
     \end{equation}
The above condition trivially follows from  $\diff \Xi^I- \tilde{{\bf H}}^I$ being everywhere zero. In particular the  $(2,1)$, $(1,2)$, $(0,3)$ components of $\diff \Xi^I- \tilde{{\bf H}}^I$ vanish and thus we can define on them the action of $\mathbb{P}_\pm$:
     \begin{equation}
\mathbb{P}_+\left(\diff \Xi^I+ Z^I\right)=0\,.\label{eeq32}
\end{equation}
In general we can write, using  \eqref{eeq42} and \eqref{eeq52}, the following relations:
\begin{equation}
    \diff\Xi^I+Z^I=\mathbb{P}_-(\tilde{H}^I)\,.\label{eqXi}
\end{equation}
Using \eqref{eeq2} and \eqref{eqXi} we find
\begin{equation}
     \mathbb{P}_+\left(\diff B^I+Z^I\right)=0\,,\,\,\, \mathbb{P}_+(\diff \tilde{P}^I)=0\,,
\end{equation}
where, as usual, we have defined $B^I\equiv (P^I+\Xi^I)/2,\,\tilde{P}^I\equiv (P^I-\Xi^I)/2$. In the above equations the action of $\mathbb{P}_+$ is well defined being both $\diff B^I+Z^I$ and $\diff \tilde{P}^I$ $(3,0)$-forms. The last of the above equations implies that $\tilde{P}^I$ is a free field. Finally, from \eqref{eeq4} and \eqref{eeq2} we find an expression for the supercovariant field strengths of $B^I$:
\begin{equation}
    H^I=\diff B^I+Z^I=\mathbb{P}_-\,(H^I)=\frac{1}{2}\left(
    \hat{H}_-^I+\mathbb{P}_-(\tilde{H}^I)\right)\,.\label{HHtHh}
\end{equation}
Let us now consider the equations for the other fields $\Phi$:
\begin{align}
      \delta_\Phi\hat{\mathcal{L}}&=-\,\delta_\Phi Z_I\wedge \left[\hat{{ H}}^I_-+\tilde{{\bf H}}^I\right]+\delta_\Phi \mathcal{L}_i\,.
\end{align}
Using Eqs. \eqref{HHtHh} and \eqref{eeq52} we can rewrite the above variation in the form:
\begin{align}
      \delta_\Phi\hat{\mathcal{L}}&=-\,\delta_\Phi Z_I\wedge \left[2\,H^I- Z^I\right]+\delta_\Phi \mathcal{L}_i\,,\label{varestn}
\end{align}
which coincides with \eqref{varest}. Equations \eqref{HHtHh} and \eqref{varestn} imply that the Euler Lagrange equations derived from $\hat{\mathcal{L}}$ are equivalent, as far as the physical sector  of the theory (consisting of $B_I,\,\phi_{AB}^I,\,\lambda_A^I$) is concerned, with those obtained from ${\mathcal{L}}$, so that condition 1 is satisfied. Once restricted to spacetime, $\hat{\mathcal{L}}$ reduces to $\tilde{\mathcal{L}}'$ (though with $Y^I=0$\footnote{We emphasize here that  the general construction discussed in the present subsection can, in principle, be applied also to rigid supersymmetric, interacting theories in different dimensions in which the forms $Z_I$ have a spacetime component $Y^I$.}) which is equivalent to Sen's spacetime description of the same model. This implies that also condition 2 is fulfilled.\par
Let us comment on the nonphysical sector which decouples from the other fields and which consists of the free fields $\tilde{P}^I$. From Eqs.  \eqref{eeq42} and \eqref{eeq52} we find
\begin{equation}
  \diff \tilde{P}^I=H^{\prime I}_-\equiv \frac{1}{2}\left(
    \hat{H}_-^I-\mathbb{P}_-(\tilde{H}^I)\right)\,.
\end{equation}
The above equations imply that $\tilde{P}^I$ are singlets with respect to supersymmetry transformations on spacetime:
$$\left.\delta_{\epsilon} \tilde{P}^I\right\vert_{\theta=0=\diff \theta}=\iota_{\epsilon}H^{\prime I}_-=\,0\,.$$
This is consistent with the analysis of \cite{Lambert:2019diy} where it was found that the free $2$-form is a singlet under supersymmetry.\par
Supersymmetry of the Lagrangian on spacetime is easily verified by 
restricting $\hat{\mathcal{L}}$ to spacetime and then using the relations \eqref{QH2} to reduce it to the Lagrangian $\tilde{\mathcal{L}}$ in Eq. \eqref{Lst1} (with $Y^I=0$). The latter is equivalent to the free Lagrangian discussed in \cite{Lambert:2019diy}. The supersymmetry transformation of $Q_-^I$ can be deduced from Eqs. \eqref{QH2} and \eqref{HHtHh}:
\begin{equation}
  \delta_\epsilon  Q_-^I=\delta_\epsilon\hat{H}^I_-+ \mathbb{P}_-(\delta_\epsilon\tilde{H}^{{I}})=2\, \mathbb{P}_-(\delta_\epsilon H^I)=\frac{1}{8}\,\mathbb{C}^{AB }\left(\overline{\epsilon}_A\Gamma_{ab}\partial_c\lambda_B^I-\frac{1}{6}\,\epsilon_{abcdef}\,\overline{\epsilon}_A\Gamma^{de}\partial^f\lambda_B^I\right)\,V^a\wedge V^b\wedge V^c\,,
\end{equation}
where 
$$\delta_\epsilon H_{abc}^{{I}}=\overline {\epsilon}_A  \nabla^A H_{abc}^{{I}}=\frac{1}{8}\,\mathbb{C}^{AB}\,\overline {\epsilon}_A\Gamma_{[ab}\partial_{c]}\lambda_B^I\,.$$
As for the $P^I$ fields we have, using $\delta_{\epsilon} \tilde{P}^I=0$ and Eqs. \eqref{susytrasfst}, that
\begin{equation}
  \delta_{\epsilon} P^I_{\mu\nu}  =\delta_{\epsilon} B^I_{\mu\nu}+\delta_{\epsilon} \tilde{P}^I_{\mu\nu}=\delta_{\epsilon} B^I_{\mu\nu}=\frac{1}{4}\,\mathbb{C}^{AB}\overline {\epsilon}_A\Gamma_{\mu\nu}\lambda_B^I\,.
\end{equation}
The supersymmetry variations of $\lambda_A^I$ and $\phi_{AB}^I$ are given in \eqref{susytrasfst}.\par
The fact that $\tilde{P}^I$ do not participate in the supersymmetric picture (being supersymmetry singlets) was to be expected since, in the presence of these fields, the on-shell matching of bosonic and fermionic degrees of freedom does not hold. Related to this is the failure of an ordinary rheonomic description for $\tilde{P}^I$. One could try to derive a consistent supersymmetric description of these fields by resorting to a form of nonlinear supersymmetry. Such a construction would however apply to an unphysical sector which decouples from the physical one and therefore we shall refrain from further dwelling on this issue in the present work, leaving this analysis to a future investigation.  \\
{As a final remark, let us notice that this first-order superspace description cannot be turned into a second-order one as for bosonic theories, because this would require the notion of the Hodge dual in superspace, which is only defined in the integral forms framework.}\par
We have thus put forward a consistent proposal for a superspace extension of Sen's prescription.

\section{Towards Full Description with Integral Forms}\label{Towards full description with integral forms}

In \eqref{susylag} we introduced the rheonomic Lagrangian as a $(6|0)$-superform $\mathcal{L}^{(6|0)} \in \Omega^{(6|0)} \left( \mathcal{M}^{(6|16)} \right)$. The spacetime manifold $\mathcal{M}^{(6)} \equiv \mathcal{M}^{(6|16)}_{red}$ coincides with the  \emph{reduced manifold} (or base manifold) and we denote with $i$ the embedding map
\begin{equation}\label{TFDIFA}
	i : \mathcal{M}^{(6)} \to \mathcal{M}^{(6|16)} \ .
\end{equation}
Vice versa, we can dualize \eqref{TFDIFA} to study the pullback of functions from the supermanifold to the reduced one, or, in general, of forms from the supermanifold to its base as
\begin{equation}\label{TFDIFB}
	i^* : \Omega^{(6|0)} \left( \mathcal{M}^{(6|16)} \right) \to \Omega^{(6)} \left( \mathcal{M}^{(6)} \right) \ ,
\end{equation}
so that we obtain a \emph{top form} on the base manifold which can be consistently integrated to define an action:
\begin{equation}\label{TFDIFC}
	S = \int_{\mathcal{M}^{(6)} \hookrightarrow \mathcal{M}^{(6|16)}} i^* \mathcal{L}^{(6|0)}.
\end{equation}

Now, we can then lift the Lagrangian to be a top form on the supermanifold by means of what is known in supergeometry as a \emph{picture changing operator} (PCO) $\mathbb{Y}^{(0|6)}$; the latter maps superforms into top forms, which are knowns as {\it integral forms}. The PCO is the \emph{Poincar\'e dual} of the embedding \eqref{TFDIFA} and it can be realized as a multiplicative operator which localizes on the reduced manifold. For example, we can write the trivial embedding
\begin{eqnarray}
	\label{TFDIFD} i : \mathcal{M}^{(6)} &\to& \mathcal{M}^{(6|16)} \\
	\nonumber \left( x_0 , \ldots , x_5 \right) &\mapsto& \left( x_0 , \ldots , x_5 , 0 , \ldots , 0 \right) \ ,
\end{eqnarray}
which corresponds to a PCO that projects on the locus $\theta^\alpha = 0 = \diff \theta^\alpha , \forall \alpha = 0, \ldots , 16$. Namely, we have
\begin{equation}\label{TFDIFE}
    \mathbb{Y}^{(0|16)}_{\rm{s.t.}} = \theta^1 \ldots \theta^{16} \delta \left( \diff \theta^1 \right) \wedge \ldots \wedge \delta \left( \diff \theta^{16} \right) \ ,
\end{equation}
where the subscript "s.t." indicates that \eqref{TFDIFE} projects on the spacetime. The action corresponding to the trivial embedding \eqref{TFDIFD} is then written as
\begin{equation}\label{TFDIFF}
    S = \int_{\mathcal{M}^{(6|16)}} \mathcal{L}^{(6|0)} \wedge \mathbb{Y}^{(0|16)}_{\rm{s.t.}} = \int_{\mathcal{M}^{(6)}} \mathcal{L}^{\rm{s.t.}} \ ,
\end{equation}
where $\mathcal{L}^{\rm{s.t.}}$ was introduced in \eqref{Lagrangianspacetime} and we are left with the integration on the base (bosonic) manifold.
$\mathbb{Y}^{(0|16)}_{\rm{s.t.}}$, as a Poincar\'e dual, is a cohomology representative (with respect to the de Rham differential) living in $H^{(0|16)} \left( \mathcal{M}^{(6|16)} , \diff \right)$. Changing the representative corresponds to the choice of different embeddings of the reduced manifold and, dually, it corresponds to adding $\diff$-exact terms to the PCO:
\begin{equation}\label{TFDIFG}
    \mathbb{Y}^{(0|16)}_{\rm{s.t.}} \mapsto \mathbb{Y}^{(0|16)} = \mathbb{Y}^{(0|16)}_{\rm{s.t.}} + \diff \Sigma^{(-1|16)} \ ,
\end{equation}
where we consider negative-degree integral forms because of the unboudedness of the integral form complex (see, e.g., Appendix \ref{integral}). 

In general, the action will be independent of the choice of representative if {$\mathcal{L}$} is closed: given two PCOs $\mathbb{Y}^{(0|16)}$ and $\mathbb{Y}'^{(0|16)}$ s.t. $\mathbb{Y}^{(0|16)} - \mathbb{Y}'^{(0|16)} = \diff \Sigma^{(-1|16)}$, we have
\begin{align}
    \nonumber S' &= \int_{\mathcal{M}^{(6|16)}} \mathcal{L}^{(6|0)} \wedge \mathbb{Y}'^{(0|16)} = \int_{\mathcal{M}^{(6|16)}} \mathcal{L}^{(6|0)} \wedge \left( \mathbb{Y}^{(0|16)} + \diff \Sigma^{(-1|16)} \right) \\
    \label{TFDIFH} &=\int_{\mathcal{M}^{(6|16)}} \mathcal{L}^{(6|0)} \wedge \mathbb{Y}^{(0|16)} - \int_{\mathcal{M}^{(6|16)}} \diff \mathcal{L}^{(6|0)} \wedge \Sigma^{(-1|16)} + \rm{b.t.} \nonumber\\
    &= S - \int_{\mathcal{M}^{(6|16)}} \diff \mathcal{L}^{(6|0)} \wedge \Sigma^{(-1|16)} + \rm{b.t.} \ ,
\end{align}
where with ``b.t.'' we denote boundary terms. If we neglect them, we immediately see that $S = S'$ if the Lagrangian is $\diff$-closed. In particular, this would mean that the action is independent of the embedding of the spacetime in the superspace. However, the closure of the Lagrangian is guaranteed only in few known cases, in particular, when it is possible to add auxiliary fields that guarantee off-shell invariance of the Lagrangian. In the case of \eqref{susylag}, it is possible to show that the Lagrangian is not closed, hence different choices of embedding give rise to different actions and, in particular, to a different number of degrees of freedom.

Alongside, the analysis of the free differential algebra associated to this model seems to suggest that it is not possible to add fields to the theory s.t. we can match (off-shell) degrees of freedom, so it seems impossible to derive a consistent closed Lagrangian. However this is not the topic of this article and will be discussed elsewhere. 

The previous argument shows that the Euler-Lagrange derived from the Lagrangian \eqref{susylag} do not coincide with the equations of motion coming from a variational principle of an action, as they do not keep track of the embedding. In other words, given an action formally written as
\begin{equation}\label{TFDIFI}
	S = \int_{\mathcal{M}^{(D|N)}} \mathcal{L}^{(D|0)} \left( \phi \right) \wedge \mathbb{Y}^{(0|N)} \ ,
\end{equation}
where we generically denote by $\phi$ the fields (eventually, forms) contained in the Lagrangian, the variational principle gives rise to \emph{constrained} equations of motion:
\begin{equation}\label{TFDIFJ}
	\delta_\phi S = 0 \ \implies \ \delta_\phi \mathcal{L}^{(D|0)} \left( \phi \right) \wedge \mathbb{Y}^{(0|N)} = 0 \ .
\end{equation}
The fact that different choices of PCO reflect different degrees of freedom of the theory (when the Lagrangian is not closed) is a consequence of the kernel of the PCO (which reflects with the kernel of the pull-back $i^*$) on $\Omega^{(D|0)} \left( \mathcal{M}^{(D|N)} \right)$, which is always nonempty.

In order to derive the self-duality condition from a superspace action, we will then need to implement Sen's principle on an action integrated on a supermanifold. In \cite{Cremonini:2020skt} the authors have shown in the easier context of the chiral boson that this corresponds to coupling the theory to an external self-dual form (actually, a pseudoform); in particular, this self-dual form needs to be coupled to the $3$-form $H^I$ and make it inherit on-shell self-duality. We will have the new action written as
\begin{equation}\label{TFDIFK}
	\tilde{S} = \int_{\mathcal{M}^{(6|16)}} \left[ \mathcal{L}^{(6|0)} \left( \phi \right) \wedge \mathbb{Y}^{(0|16)} + H^I \wedge \tilde{\mathbb{Y}}^{(0|8)} \wedge Q^{(3|8)}_I \right] \ ,
\end{equation}
where $Q^{(3|8)}_I = \star Q^{(3|8)}_I$ is the self-dual external pseudoform and $\tilde{\mathbb{Y}}^{(0|8)}$ is a half-PCO at picture equal to eight which is half of the maximal picture number, needed to lift the $(6|8)$-form $H^I \wedge Q^{(3|8)}_I$ to an integral form and ``$\star$'' is the Hodge operator on supermanifolds defined in Appendix \ref{integral}. 

\subsection{Changing the PCO}
\label{cPCO}

In order to prepare the stage for a subsequent analysis, we sketch here two alternative PCOs and 
show how the computation can be performed using the rheonomic Lagrangian \eqref{susylag}. This will be crucial to show that different embeddings pick up different terms from the Lagrangian which should contain all needed information, but with a different degree of manifest supersymmetry (in Appendix \ref{integral} some details are given). In particular, the 
amount of explicit supersymmetry is related to the number of explicit $\theta$'s in the PCO \eqref{OPCOSE}. 

We now discuss the following two examples of PCO's: the first one, which has 11 naked $\theta$'s, can be written as
\begin{eqnarray}
\label{pco1A}
\mathbb{Y}_{11} = (\epsilon \theta^{11})_{\alpha_1 A_1 \dots \alpha_5 A_5} 
(V^{a_1} \Gamma_{a_1} \iota)^{\alpha_1 A_1} \dots (V^{a_5} \Gamma_{a_5} \iota)^{\alpha_5 A_5} \delta^{16}(\psi) \,,
\end{eqnarray}
where $\iota_{\alpha A} \psi^{\beta B} = \delta_\alpha^\beta \delta_A^B$ and $\epsilon$ denotes a collection of invariant tensors of $SO(1,5)$ and $\mathbb{C}_{AB}$ to reproduce 
the Levi-Civita tensor $\epsilon$ in the 16-dimensional spinorial space.\\ 
If we multiply the rheonomic Lagrangian $\mathcal{L}^{(6|0)}$ by $\mathbb{Y}_{11}$, we select only one term 
\begin{eqnarray}
\label{pco1B}\hspace{-2cm}
&&\mathcal{L}^{(6|0)} \wedge \mathbb{Y}_{11} = \nonumber \\
&=& 
 \frac{\ii}{2} \Big( H^I \phi_{IAB}\overline\psi^A\Gamma_a\psi^B V^a \Big) 
 \wedge(\epsilon \theta^{11})_{\alpha_1 A_1 \dots \alpha_5 A_5} 
(V^{a_1} \Gamma_{a_1} \iota)^{\alpha_1 A_1} \dots (V^{a_5} \Gamma_{a_5} \iota)^{\alpha_5 A_5}  \delta^{16}(\psi)  
\nonumber \\
&=&  \alpha'  \phi_{I AB} \left( 
\mathbb{C}^{A_1 A} \mathbb{C}^{A_2 B} \mathbb{C}^{A_3 B_1} \dots \mathbb{C}^{A_5 B_3}  \right) 
\Big( \epsilon \theta^{11} T \iota^3 H^I\Big)_{A_1 \dots A_5 B_1 \dots B_3}  V^6 \delta^{16}(\psi) \,,
\end{eqnarray}
where $\alpha'$ is a suitable coefficient and
\begin{eqnarray}
\label{pco1BA}
\Big( \epsilon \theta^{11} T \iota^3 H^I\Big)_{A_1 \dots A_5 B_1 \dots B_3} &=& 
T^{\alpha_1 \dots \alpha_5 \beta_1 \dots \beta_3} 
(\epsilon \theta^{11})_{\alpha_1 A_1 \dots \alpha_5 A_5} 
H^I_{\beta_1 \dots \beta_3 B_1 B_2 B_3}  \nonumber \\ 
T^{\alpha_1 \dots \alpha_5, \beta_1 \dots \beta_3} &=& 
\epsilon^{a a_1 \dots a_5} (\Gamma_{a})_{\beta_4 \beta_5}  
(\Gamma_{a_1})^{\alpha_1 \beta_1} \dots (\Gamma_{a_5})^{\alpha_5 \beta_5} 
\nonumber \\
H^I_{\alpha\beta\gamma  ABC} &=& \iota_{\alpha A} \iota_{\beta B} \iota_{\gamma C} H^I \,.
\end{eqnarray}
The integration on the supermanifold leads to 
\begin{eqnarray}
\label{pco1C}
S_{11} &=& \int \mathcal{L}^{(6|0)} \wedge \mathbb{Y}_{11}  \nonumber \\
& =& \alpha' \left( 
\mathbb{C}^{A_1 A} \mathbb{C}^{A_2 B} \mathbb{C}^{A_3 B_1} \dots \mathbb{C}^{A_5 B_3}  \right) 
\int_{x,\theta}  
 \phi_{I AB} 
\Big( \epsilon \theta^{11} T \iota^3 H^I\Big)_{A_1 \dots A_5 B_1 \dots B_3} \nonumber \\
&=& \alpha' \left( 
\mathbb{C}^{A_1 A} \mathbb{C}^{A_2 B} \mathbb{C}^{A_3 B_1} \dots \mathbb{C}^{A_5 B_3}  \right) 
\int_x \Big(D^5  (\phi_{I AB} T \iota^3 H^I)\Big |_{\theta=0}\Big)_{A_1 \dots A_5 B_1 \dots B_3} \,,
\end{eqnarray}
where $D^5$ is the product of five superderivatives and there is only a single invariant spinorial contraction among the tensors $D^5$, $T$ and $\iota^3 H^I$.  The resulting integral over the bosonic coordinates produces a component action, as the five-order derivative $D^5$ acting on a bilinear term yields six terms of the form $D^p \phi_{IAB} D^{5-p} (\iota^3 H^I)$ with $p=0,\dots, 5$. It will be a matter of subsequent work to explore the complete component expansion of the 
action \eqref{pco1C}.

The second PCO to be discussed is the following: 
\begin{eqnarray}
\label{pco1D}
\mathbb{Y}_{14} = (\epsilon \theta^{14})_{\alpha_1 A_1\alpha_2 A_2} 
(V^{a_1} \Gamma_{a_1} \iota)^{\alpha_1 A_1}  (V^{a_2} \Gamma_{a_2} \iota)^{\alpha_2 A_2} \delta^{16}(\psi) \,.
\end{eqnarray}
Inserting it into the action, it will fish for terms with at most four explicit $V$'s and two $\psi$'s. This PCO would again lead to an action but pick some different terms as compared to \eqref{pco1C}. {In particular, it will extract the terms with $V^4 \psi^2$ from \eqref{susylag}, which are directly related to the self-duality constraint, as seen from the rheonomic equations in full superspace. In addition, the number of naked $\theta$s implies that the calculation of the Berezin integral involves only two superderivatives. Note that, compared to \eqref{pco1A} this new PCO selects different terms in the rheonomic Lagrangian. Actions written using different PCOs differ in the amount of manifest supersymmetry.} The complete expression will be presented in future work.

\section{Conclusions and outlook}

In this work, we set the basis for the complete construction of an action for noninteracting tensor multiplets {living on a stack of} M5-branes in superspace. As explained in the text, the construction amounts to deriving a {\it rheonomic} Lagrangian reproducing superspace parametrizations, the equations of motion and ready to be integrated on the full supermanifold. \\
{In the first four sections we obtain an important preliminary result in this sense, by first constructing a rheonomic 6-superform Lagrangian yielding, in superspace, all the dynamical equations, including the anti-self-duality constraint on the $3$-form field strengths. We further propose a first-order formulation of Sen's Lagrangian on spacetime and its superspace extension, which yields, on the one hand, all the rheonomic constraints on the physical fields in superspace and, on the other hand,  upon restriction to spacetime, all the dynamical equations, which include the anti-self-duality constraint on the $3$-form field strengths.} \\
In the last section, we discuss the relevant steps for the construction of an action principle in superspace through the use of integral forms and we illustrate two examples.\\

Let us conclude with some remarks. It is shown that the Lagrangian presented in Eq. \eqref{susylag} 
encodes the information about the tensor multiplet in a very compact and effective way. It is the starting point 
for a complete analysis in superspace language and for the coupling to supergravity. In addition, it would be interesting to make contact with the constructions of \cite{Galperin:2001seg,Howe:1985ar,Sokatchev:1988aa,Ferrara:2000xg,Howe:2000nq,Buchbinder:2017ozh,Buchbinder:2018bhs}, involving harmonic and pure spinor superspaces \cite{Grassi:2004xc}, which is left to future publications.
Finally, the complete Sen's mechanism for any choice of PCO will be studied deeply.  

\section*{Acknowledgments}
We thank L. Castellani, R. Catenacci, and B. L. Cerchiai for useful discussions.\\
This work was supported by the Grant Agency of the Czech Republic under the Grants No. EXPRO 20-25775X and No. EXPRO 19-28628X. \\
L. R. would like to thank the Department of Applied Science and Technology of the Polytechnic of Turin and the INFN for financial support.

\appendix

\section{Useful Formulas and Conventions}\label{conv}

We work with a mostly minus spacetime signature
$\eta_{\mu\nu}= \mathrm{diag}\left(+,-,\cdots,-\right)$.
Moreover, we adopt the following conventions:
\begin{itemize}
    \item $\epsilon_{0\ldots 5}=-\epsilon^{0\ldots 5}=1$,
    \item $\epsilon_{\mu_1\ldots\mu_k\nu_1\ldots\nu_{6-k}}\epsilon^{\mu_1\ldots\mu_k\rho_1\ldots\rho_{6-k}}=-\, k!\,\delta^{\rho_1\ldots\rho_{6-k}}_{\nu_1\ldots\nu_{6-k}}$,
    \item $\delta^{1\ldots 6-k}_{1\ldots6-k}=1$,
    \item $\diff x^{\mu_1}\wedge\ldots\wedge\diff x^{\mu_6}=-\epsilon^{\mu_1\ldots\mu_6}\diff x^{0}\wedge\ldots\wedge\diff x^{5}$,
    \item ${{}^*}_g\omega=\frac{1}{(6-k)!}\underbrace{\bigg(\frac{\sqrt{g}}{k!}\epsilon_{\mu_1\ldots\mu_k\rho_1\ldots\rho_{6-k}}\omega^{\mu_1\ldots\mu_k}\bigg)}_{({{}^*}\omega)_{\rho_1\ldots\rho_{6-k}}}\diff x^{\rho_1}\wedge\ldots\wedge\diff x^{\rho_{6-k}}$,
    \item ${{}^*}{{}^*}\omega=(-1)^{k(6-k)+1}\omega$,
    \item $\omega\wedge{{}^*}\eta=\eta\wedge{{}^*}\omega$,
    \item $(\omega,\eta)=\int\omega\wedge{{}^*}\eta$,
    \item $(\omega,\eta)=(\eta,\omega)$,
    \item $({{}^*}\omega,{{}^*}\eta)=-(\omega,\eta)$,
\end{itemize}
and we use
\begin{equation}
    \Omega^{(6)}\equiv -\frac{1}{6!}V^{abcdef}\epsilon_{abcdef}= \diff^6 x \,.
\end{equation}
For traceless antisymmetrizations in ${\rm USp}(n)$ have
\begin{equation}
    V_{[A}W_{BC]_0}= V_{[A}W_{BC]}-\frac{2}{n-2}\,\mathbb{C}_{[AB}\,W_{C]E}\,V_D\mathbb{C}^{ED}\,,
\end{equation}
where $W_{AB}$ is antisymmetric traceless. From this, setting $n=4$, we find
\begin{equation}
    V_{[A} W_{B]_0 C}=-\frac{1}{2}\,V_C\,W_{AB}+ \mathbb{C}_{C[A} W_{B]_0 E} V_D\mathbb{C}^{ED}\,.
\end{equation}

\subsection{Conventions on gamma matrices and spinors}

Our convention for the spinorial derivative is
\begin{equation}
    \overline{\psi}^A \nabla_A (\ldots) \,.
\end{equation}
The gravitino $1$-form is antichiral,
\begin{equation}
    \Psi^A = - \Gamma_7 \Psi^A \,,
\end{equation}
while the spinors $\lambda^I_A$ are chiral,
\begin{equation}
    \lambda^I_A = + \Gamma_7 \lambda^I_A \,.
\end{equation}
Besides, we have
\begin{equation}
    \Psi^A = \mathbb{C}^{AB} C \overline{\psi}^t_B \,, \quad \lambda^I_A = - \mathbb{C}_{AB} C (\overline{\lambda}^{I B})^t \,.    
\end{equation}
The 6-dimensional gamma matrices are constructed as follows:
\begin{align}
    \gamma^{\underline{a}}&=\{\sigma_1\otimes \mathbb 1_{2\times2},\ \ii\sigma_2\otimes\sigma_1,\ \ii\sigma_2\otimes\sigma_2,\ \ii\sigma_2\otimes\sigma_3\} \,, \qquad \underline{a}=0,1,2,3 \,, \\
    \Gamma^a&=\{\gamma^{\underline{a}}\otimes\sigma_1,\ \mathbb 1_{4\times4}\otimes\ii\sigma_3,\ \mathbb 1_{4\times4}\otimes\ii\sigma_2\} \,, \qquad a=0,\ldots,5 \,, \\
    \Gamma_7&=\Gamma^0\Gamma^1\Gamma^2\Gamma^3\Gamma^4\Gamma^5 \,, \quad (\Gamma_7)^2 = \mathbb{1} \,. 
\end{align}
The charge conjugation matrix is given by $C=\Gamma^1\Gamma^3\Gamma^5$ and satisfies
\begin{align}
    C&=C^t \,, \\
    C^2&=\mathbb 1_{8\times8} \,, \\
    (\Gamma_a)^t&=-C^{-1}\Gamma_aC \,.
\end{align}
The $C$-symmetry of gamma matrices is listed below:
\begin{itemize}
    \item symmetric: $\{C,\, C\Gamma^{ab}\Gamma_7,\, C\Gamma^{abc}\}$;
    \item antisymmetric: $\{C\Gamma_7,\, C\Gamma^{a},\, C\Gamma^{ab},\, C\Gamma^{a}\Gamma_7\}$.
\end{itemize}
The convention for raising and lowering the $\rm{USp}(4)$ indices is the following:
\begin{equation}
    V^A=\mathbb C^{AB}V_B \,,\qquad V_A=V^B\mathbb C_{BA}\,,
\end{equation}
where 
\begin{align}
    \mathbb C_{AB}=\mathbb C^{AB}=\begin{pmatrix}
   \mathbb 0_{2\times2} & \mathbb 1_{2\times2} \\
   -\mathbb 1_{2\times2} & \mathbb 0_{2\times2}
\end{pmatrix}\,,\qquad \mathbb C^{AC}\mathbb C_{CB}=-\delta^A_B \,.
\end{align}
The pseudo-Majorana condition for the gravitino can be written as
\begin{align}
    \psi^A=\mathbb C^{AB}C(\overline\psi_B)^t=\mathbb C^{AB}C\Gamma^0\psi^*_B=C\Gamma^0(\psi^A)^*
\end{align}
and can be inverted
\begin{align}
    \overline\psi_A=(\psi^B)^t\mathbb C_{BA}C=(\psi_A)^tC \,.
\end{align}
For $\lambda$ the result differs for a minus sign. \\

Let us also give the following useful Fierz Identities: 
\begin{align}
    \psi^A\overline\psi^B&=\frac{1}{4}\left[\Gamma_a\mathcal P_+(\overline\psi^B\Gamma^a\psi^A\mathbb)-\frac{1}{12}\Gamma_{abc}(\overline\psi^B\Gamma^{abc}\psi^A)\right]\,,\label{Fierz1}\\
    {\lambda^I_A\overline\psi^B}&{=-\frac14\mathcal P_+\overline\psi^B\lambda^I_A+\frac18\Gamma_{ab}\mathcal P_+\overline\psi^B\Gamma^{ab}\lambda^I_A}\,,\label{Fierz2}\\
    \mathbb{C}^{A[C}&\overline\psi_A \Gamma_a \psi^{B]_0}  \overline\psi_D \Gamma^a \psi^D=0\,, \label{4psi}\\
    \overline{\lambda}^D\Gamma^{ab}&\,\psi^A \overline{\psi}_C\,\Gamma_b \psi^C\,\mathbb{C}_{DA}- 4 \,\overline{\lambda}^{[D}\,\psi^{A]_0} \overline{\psi}_D\,\Gamma^a \psi^C\,\mathbb{C}_{AC}=0\,,\\
    \Gamma_{ab}\psi^A \overline\psi_B\Gamma^b\psi^B&=\psi^A \overline\psi_B\Gamma_a\psi^B- 4\psi^B \overline\psi_B\Gamma_a\psi^A
    = 4\mathbb{C}_{BC}\psi^B \psi^{ t[C}C\Gamma_a\psi^{A]_0} \,,
\end{align}
where ${\mathcal P_\pm}=\frac{\mathbb 1\pm\Gamma_7}{2}$.\\
Other useful relations are
\begin{eqnarray}
 \overline\psi_{[A}\lambda^I_{B]_0}&=&\overline\lambda^{IC}\psi^D \mathbb{C}_{D[A}\mathbb{C}_{B]_0C} \,, \\
  \overline\psi_A\Gamma_{ab}\lambda^I_{B}\mathbb{C}^{AB}&=&-\overline\lambda^{IA}\Gamma_{ab}\psi^B \mathbb{C}_{AB} \,.
\end{eqnarray}

\section{Integral Forms}
\label{integral}

In this appendix we collect some basic definitions and facts about integration on supermanifolds and \emph{integral forms}. For exhaustive introductions to integral forms we refer the reader to \cite{manin,Noja,Witten:2012bg}, while for their use in physics we refer to \cite{Castellani:2014goa, Castellani:2015paa, Cremonini:2020skt , Cremonini:2019aao , Cremonini:2019xco}.

Given a (smooth) supermanifold $\mathcal{M}^{(m|n)}$, the cotangent space $\mathcal{T}^*_P \mathcal{M}^{(D|N)}$ at a given point $P \in \mathcal{M}^{(D|N)}$ has both an even and an odd part, generated, in a given system of local coordinates $\left( x^i , \theta^\alpha \right), i=1,\ldots,D , \alpha=1,\ldots,N$, by the $(1|0)$-forms $\left\lbrace \diff x^i , \diff \theta^\alpha \right\rbrace$, called \emph{superforms}, which are respectively odd and even. They have the following (super)commuting properties: 
\begin{equation}\label{IFA}
	\diff x^i \wedge \diff x^j = - \diff x^j \wedge \diff x^i \,, \ \diff  \theta^\alpha \wedge \diff  \theta^\beta = \diff  \theta^\beta \wedge \diff  \theta^\alpha \,, \ \diff x^i \wedge \diff  \theta^\alpha = {-}\diff  \theta^\alpha \wedge \diff  x^i \ .
\end{equation}
A generic $(p|0)$-form is an object of the (graded)symmetric power of $\mathcal{T}^*_P \mathcal{M}^{(D|N)}$ and it locally reads as
\begin{equation}\label{IFB}
	\omega^{(p|0)} = \omega_{[i_1 \ldots i_r](\alpha_1 \ldots \alpha_s)} \left( x , \theta \right) \diff x^{i_1} \wedge \ldots \wedge \diff x^{i_r} \wedge \diff  \theta^{\alpha_1} \wedge \ldots \wedge \diff  \theta^{\alpha_s} \,, \ p=r+s \,,
\end{equation}
where the coefficients $\omega_{[i_1 \ldots i_r](\alpha_1 \ldots \alpha_s)}(x,\theta)$ are a set of superfields and the indices $a_1 \dots a_r$, $\alpha_1 \dots \alpha_s$ are antisymmetrized and symmetrized, respectively, as to satisfy \eqref{IFA}. We then immediately see that there is no notion of \emph{top form} among superforms, hence there is not the notion of a superform which could be integrated on $\mathcal{M}^{(D|N)}$. The notion analogous to the determinant bundle can be found in a different form complex, the complex of \emph{integral forms}. One can introduce the \emph{Berezinian bundle} $\mathpzc{B}er \left( \mathcal{M}^{(D|N)} \right)$, i.e., the space of objects which transform as the Berezinian (i.e., the \emph{superdeterminant}) under coordinate transformations. Integral forms are then constructed on open sets starting from this space and tensoring with (graded)symmetric powers of the parity-changed tangent space (see, e.g., \cite{manin} or the recent \cite{Noja} for a rigorous introduction to the subject). A practical and computationally powerful realization of the Berezinian and of integral forms is given in term of (formal) Dirac distributions on the cotangent space (see \cite{Belo} for these definitions and \cite{Witten:2012bg} for a complete review of the formalism); a generic $(p|N)$-integral form can be locally described as
\begin{equation}\label{IFC}
	\omega^{(p|N)} = \omega_{[i_1 \ldots i_r]}^{(\alpha_1 \ldots \alpha_s)} \left( x , \theta \right) \diff x^{i_1} \wedge \ldots \wedge \diff x^{i_r} \wedge \iota_{\alpha_1} \ldots \iota_{\alpha_s} \delta \left( \diff  \theta^1 \right) \wedge \ldots \wedge \delta \left( \diff  \theta^N \right) \,, \ p=r-s \,,
\end{equation}
and the second number of the $(p|n)$-form keeps track of the number of Dirac deltas and is called \emph{picture number} (see, e.g., \cite{FMS} for its introduction in string theory). The formal Dirac deltas satisfy the following properties:
\begin{align}\label{IFD}\nonumber
	\int_{\diff  \theta} \delta \left( \diff  \theta \right) &= 1, \quad \diff  \theta \wedge \delta \left( \diff  \theta \right) = 0, \quad \delta \Big( \diff  \theta^\alpha \Big) \wedge \delta \left( \diff  \theta^\beta \right) = - \delta \left( \diff  \theta^\beta \right) \wedge \delta \Big( \diff  \theta^\alpha \Big), \\ 
	\diff x \wedge \delta \left( \diff  \theta \right) &={+} \delta \left( \diff  \theta \right) \wedge \diff  x,\quad
	\delta \left( \lambda \diff  \theta \right) = \frac{1}{\lambda} \delta \left( \diff  \theta \right), \quad \diff  \theta \wedge \iota^p \delta \left( \diff  \theta \right) = - p \iota^{p-1} \delta \left( \diff  \theta \right) \ \ .
\end{align}
The first property defines how $\delta \left( \diff  \theta \right)$'s have to be used in order to perform form integration along the commuting directions $\diff  \theta$'s; the second property reflects the usual property of the support of the Dirac distribution; the third and fourth properties imply that $\displaystyle \left| \delta \left( \diff  \theta \right) \right| = 1 \mod 2$, i.e., $\delta \left( \diff  \theta \right)$'s are odd objects and together with the fifth property they indicate that actually these are not really distributions, but rather \emph{de Rham currents}, i.e., they define an \emph{oriented} integration; the last property amounts for the usual integration by parts of the Dirac delta.

A \virgolette top form'' then reads as
\begin{equation}\label{IFE}
	\omega_{top}^{(D|N)} \equiv \omega^{(D|N)} = \omega \left( x , \theta \right) \epsilon_{i_1 \ldots i_D} \diff x^{i_1} \wedge \ldots \wedge \diff x^{i_D} \wedge \epsilon_{\alpha_1 \ldots \alpha_N} \delta \left( \diff  \theta^{\alpha_1} \right) \wedge \ldots \wedge \delta \left( \diff  \theta^{\alpha_N} \right) \ ,
\end{equation}
where $\omega \left( x , \theta \right)$ is a superfield. Any integral form of any form degree $p$ can be obtained by acting with $D-p$ contractions on \eqref{IFE}. By changing the coordinate system, the $(1|0)$-forms $\diff x^a, \diff \theta^\alpha$ change as 
\begin{equation}\label{IFF}
\diff x^i \rightarrow E^a = E^a_i \diff x^i + E^a_\alpha \diff \theta^\alpha \qquad , \qquad \diff \theta^\alpha \rightarrow E^\mu = E^\mu_i \diff x^i + E^\mu_\alpha \diff \theta^\alpha \ ,
\end{equation}
where $E$ is the Jacobian (super)matrix of the transformation. A top form $\omega^{(m|n)}$  transforms as
\begin{eqnarray}\label{IFG}
\omega^{(D|N)} \rightarrow {\rm Ber}( E)  \, \omega(x,\theta) \epsilon_{i_1 \ldots i_D} \diff x^{i_1} \wedge \ldots \wedge \diff x^{i_D} \wedge \epsilon_{\alpha_1 \ldots \alpha_N} \delta \left( \diff  \theta^{\alpha_1} \right) \wedge \ldots \wedge \delta \left( \diff  \theta^{\alpha_N} \right) \ ,
\end{eqnarray}
where $ {\rm Ber}(E)$ is the superdeterminant of the (super)matrix $E$.

One can also consider other classes of forms, with nonmaximal and nonzero number of deltas: \emph{pseudoforms}. A general pseudoform with $q$ deltas is locally given by
\begin{equation}\label{IFH}
	\omega^{(p|q)} = \omega_{[a_1 \ldots a_r](\alpha_1 \ldots \alpha_s)[\beta_1 \ldots \beta_q]} \left( x , \theta \right) \diff x^{a_1} \wedge \ldots \wedge \diff x^{a_r} \wedge \diff  \theta^{\alpha_1} \wedge \ldots \wedge \diff  \theta^{\alpha_s} \wedge \delta^{(t_1)} \left( \diff  \theta^{\beta_1} \right) \wedge \ldots \wedge \delta^{(t_q)} \left( \diff  \theta^{\beta_q} \right) \ ,
\end{equation}
where we used the compact notation $\delta^{(i)} \left( \diff  \theta \right) \equiv \left( \iota \right)^i \delta \left( \diff  \theta \right)$. The form number is obtained as 
\begin{equation}\label{IFI}
	p = r + s - \sum_{i=1}^q t_i \ ,
\end{equation}
since the contractions carry negative form number. The two numbers $p$ and $q$ in Eq. \eqref{IFH} correspond to the \emph{form number} and the \emph{picture number}, respectively, and they range as $-\infty < p < +\infty$ and $0 \leq q \leq N$, so the picture number counts the number of delta's. If $q=0$ we have superforms, if $q=N$ we have integral forms, if $0<q<N$ we have pseudoforms. These kinds of forms are to be used for example in \eqref{TFDIFK} in order to construct objects which implement naturally the self-duality condition on supermanifolds. This is a consequence of the fact that the Hodge operator on supermanifolds changes not only the form number, but also the picture number:
\begin{equation}\label{IFJ}
	\star: \Omega^{(p|q)} \left( \mathcal{M}^{(D|N)} \right) \to \Omega^{(D-p|N-q)} \left( \mathcal{M}^{(D|N)} \right) \ .
\end{equation}
We refer the reader to \cite{Castellani:2014goa, Castellani:2015paa} for the introduction of the Hodge operator on supermanifolds. The action of the de Rham operator $\diff$ on pseudoforms is defined by the usual Leibniz rule and by the action on Dirac deltas as
\begin{equation}\label{IFK}
	\diff  \delta \left( E^\mu \right) = \left( \diff  E^\mu \right) \delta^{(1)} \left( E^\mu \right) \ .
\end{equation}

A notable example of integral form is the \emph{picture changing operator} described in Sec. \ref{Sec2}: it is a $(0|N)$-form, in the cohomology of the operator $\diff $. It is used to \virgolette lift'' a superform to an integral form by multiplication:
\begin{eqnarray}
	\nonumber \mathbb{Y}^{(0|N)} : \Omega^{(p|0)} \left( \mathcal{M}^{(D|N)} \right) &\to& \Omega^{(p|N)} \left( \mathcal{M}^{(D|N)} \right) \\
	\omega^{(p|0)} &\mapsto& \omega^{(p|N)} = \omega^{(p|0)} \wedge \mathbb{Y}^{(0|N)} \ .
\end{eqnarray}
As we discussed in Sec. \ref{Sec2}, its geometrical meaning is to keep track of the embedding of the reduced manifold in the supermanifold.

\subsection{Other PCOs}

Here we show how to construct PCOs corresponding to nontrivial embeddings. In particular, we show how to costruct PCOs which are manifestly invariant with respect to the Killing spinors. Infinitesimal transformations of the PCO's are described by Lie derivatives: given a vector field $v \in T_P \mathcal{M}^{(D|N)}$ they read
\begin{equation}\label{OPCOSB}
	\delta_v \mathbb{Y}^{(0|N)} = \mathcal{L}_v \mathbb{Y}^{(0|N)} = \left( \diff \iota_v + (-1)^{|v|} \iota_v \diff \right) \mathbb{Y}^{(0|N)} = \diff \iota_v \mathbb{Y}^{(0|N)} \ ,
\end{equation}
where the sign depends on the parity of $v$.
Then we see that $\mathbb{Y}^{(0|N)}$ is invariant by transformations induced by $v$ if and only if $\diff \iota_v \mathbb{Y}^{(0|N)} = 0$. In the present case, we will construct the vector $v$ in terms of the supercharge vector $Q_\alpha$. 
In particular, fixed a basis of $T_P \mathcal{M}^{(D|N)}$ $\left\lbrace \partial_a , D_\alpha \right\rbrace$, $a = 1, \ldots , D, \alpha = 1,\ldots N$ where $D_\alpha {=\partial_\alpha-\theta^\beta(C\Gamma^a)_{\alpha\beta}\partial_a}$ and the dual basis of $T^*_P \mathcal{M}^{(D|N)}$ $\left\lbrace V^a , \psi^\alpha \right\rbrace$, where $V^a = \diff x^a +{ \theta^\alpha (C\Gamma^a)_{\alpha \beta} \diff \theta^\beta} , \psi^\alpha = \diff \theta^\alpha$, the supercharge vector field reads
\begin{equation}\label{OPCOSC}
	Q_\alpha = \partial_\alpha + \theta^\beta (C\Gamma^a)_{\alpha \beta} \partial_a = D_\alpha + 2 \theta^\beta (C\Gamma^a)_{\alpha \beta} \partial_a \ , \ Q = \epsilon^\alpha Q_\alpha \ ,
\end{equation}
where $\epsilon^\alpha$ is a (Grassmann odd) spinor. Requiring that the PCO is invariant with respect to transformations generated by any $Q$ then means
\begin{equation}\label{OPCOSD}
	\diff \iota_Q \mathbb{Y}^{(0|N)} = 0 \ , \ \forall \epsilon^\alpha \ ,
\end{equation}
while requiring the same conditions for some choices of $\epsilon^\alpha$ would correspond to asking only for partial invariance. An example of maximally invariant PCO can be obtained from the spacetime one by performing the formal substitution {$\theta^\alpha \mapsto \theta^\alpha + l \diff x^a (\Gamma_a C)^{\alpha \beta} \iota_\beta$} 
\begin{equation}\label{OPCOSEA}
	\mathbb{Y}^{(0|N)}_{susy} = \epsilon_{\alpha_1 \ldots \alpha_N} \left( \theta^{\alpha_1} + l \diff x^{a_1} (\Gamma_{a_1} C)^{\alpha_1 \beta_1} \iota_{\beta_1} \right) \ldots \left( \theta^{\alpha_N} + l \diff x^{a_N} (\Gamma_{a_N} C)^{\alpha_N \beta_N} \iota_{\beta_N} \right) \delta \left( \psi^1 \right) \ldots \delta \left( \psi^N \right) \ ,
\end{equation}
and then determine a value of $l$ s.t. $ \delta_Q \mathbb{Y}^{(0|N)}_{susy} = 0$. The supersymmetry invariance of \eqref{OPCOSEA} can be verified by using
\begin{equation}\label{OPCOSF}
	\delta_Q \theta^\alpha = \epsilon^\alpha \,, \ \delta_Q \diff x^a = {+} \epsilon^\alpha (C\Gamma^a)_{\alpha \beta} \psi^\beta \ , \ \delta_Q \psi^\alpha = 0 \,,
\end{equation}
so that we have
\begin{align}
	\nonumber \delta_Q \mathbb{Y}^{(0|N)}_{susy} &= N \epsilon_{\alpha_1 \ldots \alpha_N} \left( \epsilon^{\alpha_1} {+} l \epsilon^\alpha (C\Gamma^{a_1})_{\alpha \beta} \psi^\beta (\Gamma_{a_1} C)^{\alpha_1 \beta_1} \iota_{\beta_1} \right) \ldots \left( \theta^{\alpha_N} + l \diff x^{a_N} (\Gamma_{a_N} C)^{\alpha_N \beta_N} \iota_{\beta_N} \right) \delta^N \left( \psi \right)   \\
	\nonumber &= N \epsilon_{\alpha_1 \ldots \alpha_n} \left( \epsilon^{\alpha_1} {-(-1)^s} l \epsilon^\alpha (C\Gamma^{a_1}\Gamma_{a_1}C)_{\alpha}^{\alpha_1}  \right) \ldots \left( \theta^{\alpha_N} + l \diff x^{a_N} (\Gamma_{a_N} C)^{\alpha_N \beta_N} \iota_{\beta_N} \right) \delta^N \left( \psi \right) = \\
	\label{OPCOSG}&= N \epsilon_{\alpha_1 \ldots \alpha_N} \left( \epsilon^{\alpha_1} {-(-1)^{s+t}} (-1)^s D l \epsilon^{\alpha_1} \right) \ldots \left( \theta^{\alpha_N} + l \diff x^{a_N} (\Gamma_{a_N} C)^{\alpha_N \beta_N} \iota_{\beta_N} \right) \delta^N \left( \psi \right) =0\ ,
\end{align}
where $\delta^N\left(\psi\right)=\delta(\psi^1) \wedge \ldots \wedge \delta(\psi^N)$ and where we have used the properties $\psi \iota \delta \left( \psi \right) = - \delta \left( \psi \right)$ and $\Gamma^a \Gamma_a = D \mathbb{1}$. The coefficient $s$ takes into account the $C$-symmetry of gamma matrices {whereas $t$ keeps track of the square of the charge conjugation matrix $C$}. We then see that if $l {=\frac{(-1)^{s+t}}{D}}$, $\mathbb{Y}^{(0|N)}_{susy}$ is invariant.

In the specific case of this paper, we have $D=6, N=16$ and the spinor indices $\alpha$ have to be split considering the $R$-symmetry. Then, the PCO in \eqref{OPCOSEA} reads
\begin{eqnarray}
    \nonumber \mathbb{Y}^{(0|16)}_{susy} &=& C_{\alpha_1 \alpha_2} \mathbb{C}_{A_2 A_3}\ldots C_{\alpha_{15} \alpha_{16}} \mathbb{C}_{A_{16} A_1} \left( \theta^{\alpha_1 A_1} + \frac{\ii}{3} \diff x^{a_1} (\Gamma_{a_1} C)^{\alpha_1 \beta_1} \iota_{\beta_1}^{A_1} \right) \ldots \\
    \label{OPCOSE} &\ldots& \left( \theta^{\alpha_{16} A_{16}} + \frac{\ii}{3} \diff x^{a_{16}} (\Gamma_{a_{16}} C)^{\alpha_{16} \beta_{16}} \iota_{\beta_{16}}^{A_{16}} \right) \delta \left( \psi^1 \right) \ldots \delta \left( \psi^{16} \right) \ ,
\end{eqnarray} 
where the factor $l = \frac{\ii}{3}$ comes from the transformation of $\diff x$ $\delta_Q \diff x^a = \frac{\ii}{2} \overline\epsilon_A\Gamma^a\psi^A$. Notice that each term of \eqref{OPCOSEA} or \eqref{OPCOSE} is closed and nonexact, namely a PCO itself. In particular, we can tune the PCO by choosing some terms from \eqref{OPCOSE} in order to maintain or cancel some terms of the rheonomic Lagrangian when restricting on the base manifold, as shown in Sec. \ref{cPCO}.

\end{document}